\documentclass[11pt,a4paper]{article}
\pdfoutput=1 

\usepackage{jheppub} 

\usepackage[T1]{fontenc}

\usepackage[latin9]{inputenc}

\newcommand\alphaS{\alpha_{\rm\scriptscriptstyle S}}
\newcommand\alphaSsqd{\alpha_{\rm\scriptscriptstyle S}^{2}}

\newcommand{\POWHEGBOX}{\textsc{Powheg Box}}
\newcommand\LHAPDF{{\textsc{Lhapdf}}}
\newcommand\CAESAR{{\textsc{Caesar}}}

\newcommand\FKS{{\textsc{Fks}}}
\newcommand\NNPDF{{\textsc{Nnpdf}}}
\newcommand\POWHEG{{\textsc{Powheg}}}
\newcommand\PYTHIAEIGHT{{\textsc{Pythia8}}}

\newcommand\MadSpin{{\textsc{MadSpin}}}
\newcommand\Madgraphfour{{\textsc{MadGraph4}}}
\newcommand\mgamc{{\textsc{MadGraph5\_aMC@NLO}}}
\newcommand\MCatNLO{{\textsc{MC@NLO}}}
\newcommand\SHERPA{{\textsc{Sherpa}}}
\newcommand\MINLO{{\textsc{Minlo}}}
\newcommand\MINLOPRIME{{\textsc{Minlo}$^{\prime}$}}
\newcommand\NLOPS{{\textsc{Nlops}}}
\newcommand\ST{{\textsc{ST}}}
\newcommand\STJ{{\textsc{STJ}}}
\newcommand\STJR{{{\textsc{STJ}}$^\star$}}
\newcommand\KR{K_{\scriptscriptstyle{R}}}
\newcommand\KF{K_{\scriptscriptstyle{F}}}

\newcommand\muR{\mu_{\scriptscriptstyle{R}}}
\newcommand\muF{\mu_{\scriptscriptstyle{F}}}

\title{M\Large{INLO} \LARGE $t$-channel single-top plus jet}
\preprint{\\\\CERN-TH/2018-124\\TUM-HEP-1140/18}

\author[a]{Stefano Carrazza,}
\author[b]{Rikkert Frederix,}
\author[a,c]{Keith Hamilton,}
\author[a,*]{Giulia Zanderighi\note[*]{On leave from Rudolf\
    Peierls Centre for Theoretical, Physics, 1 Keble Road, University\
    of Oxford, UK}}

\affiliation[a]{CERN, Theoretical Physics Department, CH-1211,\
  Geneva 23, Switzerland}
\affiliation[b]{Physik Department T31, Technische Universit{\"a}t\
  M{\"u}nchen,\\James-Franck-Str.~1, 85748 Garching, Germany}
\affiliation[c]{Department of Physics and Astronomy, University\
  College London, London, WC1E 6BT, UK}

\emailAdd{stefano.carrazza@cern.ch}
\emailAdd{rikkert.frederix@tum.de}
\emailAdd{keith.hamilton@ucl.ac.uk}
\emailAdd{giulia.zanderighi@cern.ch}

\date{Received: date / Accepted: \today}

\abstract{We present a next-to-leading order accurate simulation
	of $t$-channel single-top plus jet production matched to
	parton showers via the \POWHEG{} method. The calculation
	underlying the simulation is enhanced with a process-specific
	implementation of the multi-scale improved NLO (\MINLO{})
	method, such that it gives physical predictions all through
	phase space, including regions where the jet additional
	to the $t$-channel single-top process is unresolved. We further
	describe a tuning procedure for the \MINLO{} Sudakov form
	factor, fitting the coefficient of the first
        subleading term in its exponent using an artificial
        neural-network. The latter tuning, implemented as a straightforward
        event-by-event reweighting, renders the \MINLO{} simulation
        NLO accurate for $t$-channel single-top observables, in
        addition to those of the analogous single-top plus jet
        process.
}


\begin{document}
\maketitle
\flushbottom

\section{Introduction}
\label{sec:intro}

The top quark is the heaviest of all the known elementary particles.
Owing to the closeness of its mass to the electroweak scale, and its
large Yukawa coupling to the Higgs boson, the top quark is considered
to have a special role in, and be a sensitive probe of, both the mechanism
of electroweak symmetry breaking~\cite{Higgs:1964pj,Englert:1964et}
and physics beyond the standard model. 
The top is further unique among the quarks insofar as it predominantly
decays before hadronizing. This fact implies that one can explore the
electroweak properties of a bare quark~\cite{Bernreuther:2008ju}. 

At the LHC the production rate of top quarks is large. For $pp$
collisions at 13 TeV centre-of-mass energy, with a top mass of
$172.5\,\mathrm{GeV}$, the main channel, top pair-production, is
predicted to have a cross section of $832^{+40}_{-46}$ pb
\cite{Cacciari:2011hy,Czakon:2011xx,Czakon:2013goa,Czakon:2017wor}.
Hence, by the end of Run II of the LHC, later this year, 100 million
top pairs will have been produced by this mechanism. These very large
event samples facilitate precise measurements of fundamental top-quark
properties, such as its mass and its couplings to other standard model
particles.

The second largest mechanism for the production of top quarks at the
LHC is through electroweak single-top production, in which a down-type
quark --- typically the bottom quark --- is converted to a top quark
by interacting with a $W$-boson. The cross section for this process
is very close to being one third that of top-quark pair production at
the 13 TeV
LHC~\cite{Aliev:2010zk,Kant:2014oha,Kidonakis:2010ux,Kidonakis:2013zqa}.
%
Despite being a slightly less copious source of top quarks and a
difficult reaction to analyse from an experimental perspective,
single-top production is nevertheless a uniquely interesting process
to study. Perhaps most notably it provides a means to directly measure
the Cabibbo-Kobayashi-Maskawa (CKM) matrix element $|V_{tb}|$, which
is otherwise only measured
indirectly~\cite{Alwall:2006bx,Lacker:2012ek,Cao:2015doa,Alvarez:2017ybk}.
Additionally, in numerous beyond the standard model scenarios, single-top
production provides a more sensitive probe of new physics than
other processes~\cite{Tait:2000sh,Cao:2007ea,Atwood:2000tu,Drueke:2014pla,%
  Aguilar-Saavedra:2017nik,Zhang:2016omx}.
Finally, besides being an interesting process in its own right, it
is also important to have a detailed understanding of single-top
production in order to control it as a background in other standard
model analyses and new physics searches, e.g.~standard model Higgs
production in association with a $W$-boson.

Single-top production is usually classified in three separate modes,
based on the virtuality of the participating $W$-boson. The
$t$-channel mechanism, where the $W$-boson has a negative virtuality,
has the largest cross section at the LHC. This is followed by the
associated $Wt$ production mode, where the virtuality of the $W$-boson
is zero, whose cross section is three to four times smaller than that
of the $t$-channel process, for centre-of-mass energies in the range
$7-14\,\mathrm{TeV}$. The $s$-channel production mode, in which the
$W$-boson has positive virtuality, has the lowest rate of all three
single-top channels, being up to a factor of ten smaller than the $Wt$
channel at the LHC.~\footnote{At higher orders in perturbation theory
  these production modes interfere, this is discussed in detail in
  Sec.~\ref{subsec:definition}.}
In this work we concentrate on the dominant mode, namely, $t$-channel
single-top production.

Experimental analysis of $t$-channel single-top production is
particularly difficult at the LHC due to the large background
from $t\bar t$ and $W$+jets events. Even so, this process
has been measured and studied by both ATLAS~\cite{Aad:2012ux,%
  Aad:2014fwa,Aad:2015yem,Aaboud:2016ymp,Aaboud:2017pdi,Aaboud:2017aqp}
and CMS~\cite{Chatrchyan:2011vp,Chatrchyan:2012ep,Khachatryan:2014iya,%
  Khachatryan:2015dzz,Sirunyan:2016cdg,Khachatryan:2016sib,Sirunyan:2017huu}
at 7, 8 and 13 TeV. For recent reviews on experimental studies
of single-top production both at the Tevatron and LHC see
refs.~\cite{Giammanco:2015bxk,Giammanco:2017xyn}.

On the theoretical front, predictions for hadronic single-top
production processes in the framework of fixed order perturbative
QCD have also been a subject of considerable work and progress.
$t$-channel single-top production has been computed at next-to-leading
order (NLO) in QCD perturbation theory, including NLO corrections
to the top-quark decays, in refs.~\cite{Harris:2002md,Campbell:2004ch,%
  Cao:2005pq}. These calculations were carried out in the so-called
five-flavour scheme, in which the $b$-quark is treated in the massless
approximation. More recently, NLO accurate calculations were carried
out and implemented in the \textsc{MCFM} Monte Carlo package in the
four flavour scheme~\cite{Campbell:2009ss,Campbell:2009gj} --- wherein
the $b$-quark is instead treated as a massive parton --- including
spin correlations, in the zero-width approximation, for the top-quark
decays~\cite{Campbell:2012uf}.  
Ground-breaking work in the last four years has seen the accuracy of
fixed order perturbative predictions for $t$-channel single-top
production further extended to next-to-next-to-leading order
(NNLO)~\cite{Brucherseifer:2014ama,Berger:2016oht,Berger:2017zof}, in
the approximation in which one neglects $\mathcal{O}(\alphaS^2)$
colour suppressed interference terms.

Beyond fixed order perturbation theory, all orders analytic resummation
of threshold logarithms has been presented in refs.~\cite{Wang:2010ue,%
  Kidonakis:2011wy}, with transverse momentum resummation effects having
been studied in ref.~\cite{Cao:2018ntd}.
Precision Monte Carlo simulations of $t$-channel single-top production
processes have also been developed, based on the matching of NLO
calculations with parton showers (\NLOPS{}), in the
\MCatNLO{}~\cite{Frixione:2005vw,Frixione:2008yi,Frederix:2012dh},
\POWHEG{}~\cite{Alioli:2009je,Frederix:2012dh}, and
\SHERPA{}~\cite{Bothmann:2017jfv} frameworks.

Off-shell top-quark effects have also been considered at NLO, both
at fixed order~\cite{Falgari:2010sf,Falgari:2011qa} and further in
the context of NLO parton shower matched
simulations~\cite{Papanastasiou:2013dta,Jezo:2015aia,Frederix:2016rdc}.
These studies reveal such effects to be small away from kinematic
end-points. Finally, electroweak corrections to $t$-channel single-top
production have been computed and also found to be
small~\cite{Beccaria:2008av,Bardin:2010mz,Frederix:2018nkq}, 
affecting the total cross section at the sub-percent level, but
with the effects rising in regions where the kinematic invariants
associated with the process become large.

While the \NLOPS{} Monte Carlo description of single-top production
reaches a remarkable level of accuracy and sophistication, it is 
not as advanced as that afforded to other processes, such as Higgs,
$W$-, and $Z$-boson production. In particular, 
powerful methods recently developed for merging together \NLOPS{} simulations of processes
that differ only in their jet multiplicity~\cite{Alioli:2011nr,%
  Hamilton:2012np,Hoeche:2012yf,Frederix:2012ps,Platzer:2012bs,%
  Alioli:2012fc,Lonnblad:2012ix,Hamilton:2012rf,Alioli:2013hqa,%
  Bellm:2017ktr}, e.g.~Higgs and Higgs plus
jet production simulations, have not so far been applied to single-top.
With the exception of relative $\mathcal{O}(\alpha_{\scriptscriptstyle{S}}^{2})$
virtual corrections, event generators based on these methods contain
all of the same fixed order information as found in NNLO calculations,
all consistently matched to leading-log parton shower resummation and
tuned non-perturbative models.
%

In the present work, we constructed
a first \NLOPS{} simulation of $t$-channel single-top plus jet
production within the \POWHEGBOX{} framework~\cite{Alioli:2010qp,%
  Alioli:2010xd}, with matrix elements obtained from \mgamc{} and
related packages~\cite{Alwall:2014hca,Ossola:2006us,Ossola:2007ax,%
  shao,vanHameren:2010cp}. We then enhanced the underlying NLO
calculation according to the multiscale improved NLO (\MINLO{})
procedure~\cite{Hamilton:2012np}, with important but straightforward
specializations to the case at hand. Finally, we have invoked the basic
idea put forward in ref.~\cite{Frederix:2015fyz}, with substantial
refinements and extensions, to recover NLO accuracy in the lower
multiplicity $t$-channel single-top process, by approximately fitting
unknown, subleading, $\mathcal{O}(\alpha_{\scriptscriptstyle{\mathrm{S}}}^{2})$
terms in the \MINLO{} Sudakov form factor. In particular, to tune
the latter Sudakov form factor we make use of machine learning methods
in the form of an artificial neural network. 

Formally, the minimal, sufficient condition for carrying out tuning
to this effect is merely that, beforehand, the \MINLO{} $t$-channel
single-top plus jet computation must be at least LO accurate for
inclusive $t$-channel single-top production observables.
This is implicitly the case if the resummation formula underlying the
initial \MINLO{} simulation is next-to-leading-log ($\mathrm{NLL}_{\sigma}$)
accurate, as in this work.\footnote{
  $\mathrm{NLL}_{\sigma}$ resummation controls all terms in the $t$-channel
  single-top plus jet cross section 
  $\propto \frac{1}{y_{12}}\,\bar{\alpha}_{{\scriptscriptstyle \mathrm{S}}}^{n}\,\ln^{m}\frac{Q^{2}}{y_{12}}$,
  with $m=2n-1$ and $m=2n-2$, wherein $Q$ is a scale characteristic
  of the hard, underlying, $2 \rightarrow 2$ scattering, and $y_{12}$
  is the value of the distance measure in the exclusive $k_{t}$
  clustering algorithm \cite{Catani:1993hr}, where a single-top
  plus jet event is resolved as a single-top one.
  $\mathrm{NNLL}_{\sigma}$ resummation also controls $m=2n-3$ terms.} 
If the latter condition is satisfied, the desired NLO corrections to
inclusive $t$-channel single-top observables can be accounted for by
introducing a ${\mathcal{O}}(\mathrm{NNLL}_{\sigma})$ term in the
\MINLO{} Sudakov form factor, with a fitted $\mathcal{O}(1)$ coefficient.
This is a straightforward mathematical fact, that need not have
anything to do with resummation.
The tuning procedure will raise the \MINLO{} $t$-channel single-top
plus jet description of inclusive $t$-channel single-top observables
to NLO accuracy by construction. At the same time, since this is
achieved by introducing only a ${\mathcal{O}}(\mathrm{NNLL}_{\sigma})$
term in the \MINLO{} Sudakov exponent, the NLO accuracy already
in place for $t$-channel single-top plus jet production will remain
intact.

Introducing higher order terms in Sudakov form factors, to unitarize
cross sections, be they spurious or not, is not new. For example, the
\textsc{H/W/Zj}-\MINLOPRIME{} simulations of ref.~\cite{Hamilton:2012rf},
achieving the same level of accuracy that we aim for in this work,
have this property.
The \textsc{H/W/Zj}-\MINLOPRIME{} constructions eliminated
${\mathcal{O}}(\alpha_{\scriptscriptstyle{\mathrm{S}}}^{3/2})$ differences
between \textsc{H/W/Zj}-\MINLO{} predictions for inclusive observables
and conventional NLO ones, by adding ${\mathcal{O}}(\mathrm{N^3LL}_{\sigma})$,
`$B_2$', terms in the \MINLO{} Sudakov form factor.
While inclusion of the latter `$B_2$' terms led to the desired level
of fixed order accuracy, their inclusion is completely spurious from
the point of view of resummation: the resummation accuracy associated
to those simulations before and after inclusion of the `$B_2$' terms
is completely unchanged. This owes to the fact that the resummation
formula underlying those simulations is based on resumming the
\textsc{H/W/Z} transverse momentum spectrum directly in transverse momentum
space.
%
%

As already stated above, to reach our (fixed order) accuracy goals, we
are only required to control terms at the $\mathrm{NLL}_{\sigma}$
level in the \MINLO{} resummation formula prior to invoking the
tuning procedure: we do not require any information on the form,
or ingredients, of a more accurate formula for this aim.
Nevertheless, we postulate that the only difference between
the \emph{form} of our $\mathrm{NLL}_{\sigma}$ \MINLO{} resummation
formula and its $\mathrm{NNLL}_{\sigma}$ extension, merely lies in
the inclusion of a $\mathrm{NNLL}_{\sigma}$ term in the Sudakov form
factor. Thus, while it is unnecessary for achieving our desired
level of fixed order accuracy, if our mild postulate on the form
of the $\mathrm{NNLL}_{\sigma}$ resummation formula holds, when rendering
\MINLO{} $t$-channel single-top plus jet NLO accurate for inclusive
$t$-channel single-top observables, we will implicitly also improve
the $\mathrm{NLL}_\sigma$ \MINLO{} resummation towards the true
$\mathrm{NNLL}_\sigma$ result.

In section~\ref{sec:method} we present the theoretical framework, charting
the construction of our simulation: first the NLO computation, followed
by its \MINLO{} extension, and on to the tuning of the latter Sudakov
form factor. In section~\ref{sec:results} we validate the $t$-channel
single-top \MINLO{} simulation, \STJ{}, and its tuned counterpart, \STJR{},
by comparing their predictions to one another, as well as to those of the
lower multiplicity \POWHEG{} $t$-channel single-top production code,
\ST{}~\cite{Alioli:2009je}. We conclude in section~\ref{sec:conclusions}.
Finally, appendix~\ref{app:Supplementary-details-on-the-MiNLO-procedure}
provides supplementary details on the theoretical framework, while
additional numerical results are given in
appendix~\ref{app:Tune-extrapolation}, to give insight on the
robustness of the tuning in the \STJR{} simulation.


\section{Theoretical framework}
\label{sec:method}

In this section we describe the main elements of our $t$-channel
single-top plus jet simulation and their connections. In
section~\ref{subsec:definition} we give details on the precise definition
of the $t$-channel single-top process, addressing issues that arise
there due to ambiguities at $\mathcal{O}(\alphaSsqd{})$.
Section~\ref{subsec:NLOPS-construction} documents the matrix elements
used in building our initial \NLOPS{} $t$-channel single-top plus
jet simulation, and its assembly in the \POWHEGBOX{} framework.
The enhancement of the latter with a process-specific adaptation
of the multi-scale improved NLO (\MINLO{}) method is described in
section~\ref{subsec:STJ-MiNLO}. In section~\ref{subsec:Improved-STJ-MiNLO}
we describe how one can tune the Sudakov form factor in the latter
\STJ{} simulation such that it recovers NLO accurate predictions
for inclusive $t$-channel single-top production observables,
while retaining NLO accuracy for single-top plus jet ones.
Section~\ref{sec:NN} goes on to describe a concrete realization
of this method, making use of machine learning algorithms.

For brevity, throughout our work, we will refer to the inclusive
$t$-channel single-top process as \ST{}, and the $t$-channel
single-top plus jet process as \STJ{}. The abbreviation \ST{} will
be further used to denote the \POWHEG{} inclusive $t$-channel
single-top production program~\cite{Alioli:2009je}, which is NLO
accurate in the description of inclusive $t$-channel single-top
observables. Similarly, we will use \STJ{} to refer to our
\MINLO{} $t$-channel single-top plus jet simulation, NLO accurate
in the description of \STJ{} observables. The tuned counterpart
of \STJ{} is labelled \STJR{}.

\subsection{Process definition}
\label{subsec:definition}
We consider the $t$-channel single-top production process in the
five-flavour scheme, wherein, at the lowest order in perturbation
theory, a massless initial state bottom quark is converted to a top
quark through the exchange of a $t$-channel, space-like $W$-boson. The
$W$-boson is also connected to another quark line, in which an initial
state up-type quark (down-type anti-quark) from the first two
generations is converted to a final state down-type quark (up-type
anti-quark), again from the first two generations. We refer to
these quark lines as the \textit{heavy quark line} and the
\textit{light quark line}, respectively.

To NLO in perturbation theory, $\mathcal{O}(\alphaS)$, radiative
corrections to $t$-channel single-top production factorise exactly
into independent corrections to the heavy and light quark lines,
respectively. Moreover, to NLO, $t$-channel single-top production
does not interfere with other single-top production modes. On the
other hand, when considering $\mathcal{O}(\alphaS^2)$ terms,
as in NLO $t$-channel single-top plus jet production, contributions
to the cross section start to arise from interference between
radiative corrections to the heavy and light quark lines. Since the
heavy and light quark lines correspond to two different colour lines,
interference of their associated radiative corrections amounts
to an interference of colour structures. Correspondingly, such
$\mathcal{O}(\alphaS^2)$ contributions are suppressed
by at least two powers of the number of colours, $N_{c}=3$, relative
to those involving no dynamical correlation between the heavy and
light quark lines \cite{Brucherseifer:2014ama,Berger:2016oht,Berger:2017zof}.
A non-zero interference of $s$- and $t$-channel single-top production
modes is also understood to develop at $\mathcal{O}(\alphaS^2)$
\cite{Berger:2017zof}.

The goal of this work is to construct a simulation which is NLO accurate
in the description of $t$-channel single-top and $t$-channel single-top
plus jet production. For the former, $\mathcal{O}(\alphaS^2)$ terms do
not contribute to NLO. Taking the latter point together with the expectation
that the aforementioned $\mathcal{O}(\alphaS^2)$ colour suppressed
interference terms are small, we shall omit them throughout our work.
Neglecting these contributions, treating the radiative corrections to
heavy and light quark lines as being dynamically independent of one another,
is known in the literature as the \textit{structure function approximation}
\cite{Han:1992hr}. Working in this approximation is equivalent to treating
radiative corrections to the heavy and light quark lines as if they
originated from two independent copies of the QCD sector, with cross-talk
between the two only occurring indirectly, via the electroweak sector
\cite{Cacciari:2015jma}. Dropping the $\mathcal{O}(\alphaS^2)$ colour
suppressed terms simultaneously removes the problem of how to define our
process in the presence of interfering $s$- and $t$-channel contributions
at this order, since the latter have the same physical origins as the former
\cite{Berger:2017zof}. Hence, in the context of the approximation within
which we are working, the $t$-channel single-top production process is
unambiguously defined.\footnote{This is true to all orders in QCD,
  yet it fails when higher orders in the weak coupling are considered.}

\subsection{\NLOPS{} $t$-channel single-top plus jet\label{subsec:STJ-NLO}}
\label{subsec:NLOPS-construction}

NLO accurate parton shower simulations of $s$- and $t$-channel single-top
production processes have been constructed in recent years according to
the \POWHEG{} method, and they have been well used by the ATLAS and CMS
collaborations \cite{Alioli:2009je}. Our first goal in this work has been
to develop a new \NLOPS{} simulation of $t$-channel single-top plus
jet production using the \POWHEGBOX{} framework.

To this end we have obtained the relevant Born and real matrix elements
using the \Madgraphfour-\POWHEGBOX{} interface presented in
ref.~\cite{Campbell:2012am}.
In doing so we omit diagrams with $s$-channel $W$ bosons that the interface
produces by default. The latter restriction was implemented by delicate
modifications to the \Madgraphfour{} output. Although all of the correct
diagrams are generated by the interface, some of the colour factors
associated with subleading-colour contributions in the real-emission
matrix elements required manual adjustments. The \POWHEGBOX{} Born and real
matrix elements were subsequently found to yield complete point-by-point
agreement with the analogous predictions of \mgamc{}.
To remove the interference between the corrections to the light and
heavy quark lines and be consistent with the structure function
approximation (section~\ref{subsec:definition}), a semi-automatic script
was developed to update the colour matrices present in the real-emission
and colour-correlated Born matrix elements. The convergence of the
real-emission matrix elements towards the subtraction terms for
phase-space points nearing to the soft and/or collinear limits yields
a powerful check of the consistency of not only this script, but also
the complete implementation of the Born and real matrix elements.

The virtual matrix elements for our \NLOPS{} $t$-channel single-top
plus jet simulation have been obtained using the standalone version
of \textsc{MadLoop}~\cite{Hirschi:2011pa,Alwall:2014hca}. The latter
was used to generate a library which we have directly linked to our
\POWHEGBOX{} simulation code. The library contains the virtual matrix
elements and their associated integral reduction packages,
\textsc{CutTools}~\cite{Ossola:2006us,Ossola:2007ax} and
\textsc{IREGI}~\cite{shao}, as well as
\textsc{OneLOop}~\cite{vanHameren:2010cp} for the evaluation of the
one-loop scalar integrals. By employing a \mgamc{} model which only
accounts for NLO QCD corrections, the propagators entering all of
our one-loop matrix elements only contain QCD charged particles: the
$W$-boson cannot be part of the loop itself. Thus, the virtual corrections
we have generated are fully consistent with the structure function
approximation that we have based our work on (see section~\ref{subsec:definition}).

We have carefully validated our implementation of all of the above
elements in the \POWHEGBOX{}, by comparing our predictions for the
total NLO cross section, as well as key differential distributions,
at fixed order, to those of \mgamc{}. In all cases we found complete
agreement between the two codes.

While we do not consider the decays of the top quarks in this work,
spin correlations between the top production and decay processes can
be important~\cite{Mahlon:1996pn}. The latter can be accounted for
a posteriori by decaying the top quarks with the \MadSpin{}
program~\cite{Frixione:2007zp,Artoisenet:2012st}. \MadSpin{}
can parse the Les Houches event files generated by our \POWHEGBOX{} code,
simulating the decay of the top quark in each event, including all
tree-level correlations between production and decay, to yield a new
Les Houches event file wherein all tops have been decayed.

\subsection{\MINLO{}}
\label{subsec:STJ-MiNLO}
In the \POWHEG{} framework all events generated in the calculation
of the NLO cross section have a common associated underlying Born
configuration, $\Phi_{{\scriptscriptstyle \mathrm{STJ}}}=\{q_{i}\}$,
with $\{q_{i}\}$ being the corresponding set of five momenta. The
first step in the \MINLO{} procedure is to input the
$\Phi_{{\scriptscriptstyle \mathrm{STJ}}}$
configuration to the exclusive $k_{t}$ algorithm\footnote{
  \label{footnote:flavour-sensitive-kt-clustering}
  Since we aim to provide a fully exclusive simulation and we have
  access to all particle flavours, we employ a slightly modified version
  of the exclusive $k_{t}$ algorithm where we veto clusterings of two
  particles that cannot be produced by a QCD branching. This is simply
  achieved by setting the $k_{t}$ algorithm distance measure to infinity
  if any two partons that it attempts to combine cannot be associated
  with the QCD branching of a quark or gluon~\cite{Hamilton:2012np}.}
\cite{Catani:1993hr}, yielding a $bq\rightarrow tq^{\prime}$ state together
with an associated $k_{t}$-clustering scale, $\sqrt{y_{{\scriptscriptstyle 12}}}$.
Denoting the clustering operation $\mathbb{P}$, we notate the resulting
set of $2\rightarrow2$ momenta as
$\Phi=\Phi_{{\scriptscriptstyle \mathrm{ST}}}%
=%
\{p_{i}\}\equiv\mathbb{P}[\Phi_{{\scriptscriptstyle \mathrm{STJ}}}]$.

In the limit that $\sqrt{y_{{\scriptscriptstyle 12}}}$ is small relative
to any hard scales in $\Phi$, the $t$-channel single-top plus jet cross
section is dominated by large Sudakov logarithms at all orders in
perturbation theory, rendering fixed order predictions of little or no
use, depending on the extent to which the second jet is unresolved.
\MINLO{} augments the latter NLO cross section to maintain predictivity
when such regions of phase space are probed, by matching it to an
all orders summation of these large logarithms, according to the following
formula: 
\begin{equation}
  d\sigma_{{\scriptscriptstyle \mathcal{M}}}=%
  \Delta(y_{{\scriptscriptstyle 12}})\,%
  \left[\,%
    d\sigma_{{\scriptscriptstyle \mathrm{NLO}}}^{{\scriptscriptstyle \mathrm{STJ}}}%
    -%
    \left.%
    \Delta(y_{{\scriptscriptstyle 12}})%
    \right|_{\bar{\alpha}_{{\scriptscriptstyle \mathrm{S}}}}%
    \,d\sigma_{{\scriptscriptstyle \mathrm{LO}}}^{{\scriptscriptstyle \mathrm{STJ}}}\,%
    \right]\,.%
  \label{eq:sect22-dsigma-M-in-terms-of-dsigma-NLO-etc}
\end{equation}
In eq.~\eqref{eq:sect22-dsigma-M-in-terms-of-dsigma-NLO-etc}
$d\sigma_{{\scriptscriptstyle \mathrm{LO}}}^{{\scriptscriptstyle \mathrm{STJ}}}$, $d\sigma_{{\scriptscriptstyle \mathrm{NLO}}}^{{\scriptscriptstyle \mathrm{STJ}}}$,
and $d\sigma_{{\scriptscriptstyle \mathcal{M}}}$ are the LO, NLO,
and \MINLO{} cross sections, fully differential in the three-particle
phase space of the single-top plus jet Born-like terms, and the
four-particle phase space of their real emission counterparts. All
instances of the renormalization and factorization scales in
$\Delta(y_{{\scriptscriptstyle 12}})$,
$d\sigma_{{\scriptscriptstyle \mathrm{LO}}}^{{\scriptscriptstyle \mathrm{STJ}}}$
and $d\sigma_{{\scriptscriptstyle \mathrm{NLO}}}^{{\scriptscriptstyle \mathrm{STJ}}}$
have been set to $\sqrt{y_{{\scriptscriptstyle 12}}}$. The \MINLO{}
Sudakov form factor is denoted by $\Delta(y_{{\scriptscriptstyle 12}})$,
with
$\left.\Delta(y_{{\scriptscriptstyle 12}})\right|_{\bar{\alpha}_{{\scriptscriptstyle \mathrm{S}}}}$
representing the $\mathcal{O}(\bar{\alpha}_{{\scriptscriptstyle \mathrm{S}}})$
  term in its expansion, $\bar{\alpha}_{{\scriptscriptstyle \mathrm{S}}}$
being defined as 
\begin{equation}
  \bar{\alpha}_{{\scriptscriptstyle \mathrm{S}}}=%
  \frac{\alpha_{{\scriptscriptstyle \mathrm{S}}}}{2\pi}\,.%
  \label{eq:sect22-aSbar-defn}
\end{equation}

The Sudakov form factor can be written as the product of those associated
with the light-quark ($qq^{\prime}$) and the heavy quark ($bt$) colour
dipoles in the leading order single-top process,\footnote{QCD corrections
  to the $bt$ and $qq^{\prime}$ fermion lines in the single-top process
  are completely independent of one another. Potential contributions to
  the cross section due to interference of gluons emitted from the two
  different fermion lines are readily found to be proportional to traces
  of single Gell-Mann matrices. } 
\begin{equation}
  \Delta(y_{{\scriptscriptstyle 12}})=%
  \Delta_{{\scriptscriptstyle qq^{\prime}}}(y_{{\scriptscriptstyle 12}})\,%
  \Delta_{{\scriptscriptstyle bt}}(y_{{\scriptscriptstyle 12}})\,.%
  \label{eq:sect22-Delta-eq-Delta-qqprime-Delta-bt}
\end{equation}
The $qq^{\prime}$ Sudakov form factor is given by
\begin{equation}
  \ln\Delta_{{\scriptscriptstyle qq^{\prime}}}(y_{{\scriptscriptstyle 12}})\,%
  =\,%
  -%
  2 \int_{y_{12}}^{Q_{{\scriptscriptstyle qq^{\prime}}}^{2}}\,%
  \frac{dq^{2}}{q^{2}}\,%
  \bar{\alpha}_{{\scriptscriptstyle \mathrm{S}}}^{{\scriptscriptstyle \mathrm{CMW}}}\,%
  C_{F}\,%
  \left[\,\ln\frac{Q_{{\scriptscriptstyle qq^{\prime}}}^{2}}{q^{2}}-\frac{3}{2}\,%
    \right]\,%
  ,%
  \qquad%
  Q_{{\scriptscriptstyle qq^{\prime}}}^{2}%
  =%
  2p_{q}.p_{q^{\prime}}\,%
  ,%
  \label{eq:sect22-Delta-qqprime}
\end{equation}
while the $bt$ Sudakov form factor carries additional terms in the
integrand which vanish in the limit $m_{t}\rightarrow0$, 
\begin{align}
  \ln\Delta_{{\scriptscriptstyle bt}}(y_{{\scriptscriptstyle 12}})\,%
  =%
  \, & -2 \int_{y_{12}}^{Q_{{\scriptscriptstyle bt}}^{2}}\,%
  \frac{dq^{2}}{q^{2}}\,%
  \bar{\alpha}_{{\scriptscriptstyle \mathrm{S}}}^{{\scriptscriptstyle \mathrm{CMW}}}\,%
  C_{F}\,%
  \left[\,%
    \ln\frac{Q_{{\scriptscriptstyle bt}}^{2}}{q^{2}}-\frac{3}{2}\,%
    \right]%
\nonumber \\
  \, & -\int_{y_{12}}^{Q_{{\scriptscriptstyle bt}}^{2}}\,%
  \frac{dq^{2}}{q^{2}}\,%
  \bar{\alpha}_{{\scriptscriptstyle \mathrm{S}}}^{{\scriptscriptstyle \mathrm{CMW}}}\,%
  C_{F}\,%
  \left[\,%
    \frac{1}{2}%
    -\frac{q}{m_{t}}\arctan\frac{m_{t}}{q}%
    -\frac{2m_{t}^{2}-q^{2}}{2m_{t}^{2}}\,%
    \ln\frac{m_{t}^{2}+q^{2}}{q^{2}}\,%
    \right]\,%
  ,\label{eq:sect22-Delta-bt-ii}
\end{align}
with 
\begin{equation}
  Q_{{\scriptscriptstyle bt}}^{2}\,%
  =\,%
  2p_{b}.p_{t}\,.%
  \label{eq:sect22-Qbt-eq-two-pb-dot-pt}
\end{equation}
The strong coupling is evaluated in the Bremsstrahlung (CMW) scheme
\cite{Catani:1990rr} with $q$ as its argument in
eqs.~(\ref{eq:sect22-Delta-qqprime}-\ref{eq:sect22-Delta-bt-ii}):
\begin{equation}
  \bar{\alpha}_{{\scriptscriptstyle \mathrm{S}}}^{{\scriptscriptstyle \mathrm{CMW}}}
  =
  \bar{\alpha}_{{\scriptscriptstyle \mathrm{S}}}\,
  \left[\,1+\bar{\alpha}_{{\scriptscriptstyle \mathrm{S}}}K\,\right]\,,
  \qquad K\,=\,\left[\,\frac{67}{18}-\frac{\pi^{2}}{6}\,\right]\,C_{A}-%
  \frac{10}{9}\,n_{f}T_{R}\,.
\end{equation}
We must stress that our use of
$\bar{\alpha}_{{\scriptscriptstyle \mathrm{S}}}^{{\scriptscriptstyle \mathrm{CMW}}}$
here is superfluous in the context of our work, since we do not
claim to fully control terms at that order in the Sudakov form factor
anyway ($\mathrm{NNLL}_{\sigma}$). We note its use merely to accurately
document the implementation. For all of the following discussions
its presence is irrelevant.\footnote{
  Our final numerical results actually suggest that the basic \STJ{}
  simulation would, by itself, better reproduce NLO inclusive $t$-channel
  single-top production predictions, were it to have less of the
  additional Sudakov suppression that the CMW scheme brings.
}

While the $m_{t}\rightarrow0$ limit of the Sudakov form factor here
follows directly from the \CAESAR{} formalism\footnote{With due care
  to include the soft-wide-angle term for colour dipoles prescribed by
  that framework ($S_{1}$).}  \cite{Banfi:2004yd}, we have assembled
the form factor with the full top mass dependence using the
resummation framework of ref.~\cite{Bonciani:2003nt}, elaborated on in
ref.~\cite{Frixione:2007vw}.  We further derived
$\Delta_{{\scriptscriptstyle qq^{\prime}}}$ and $\Delta_{bt}$,
independently, by an explicit $\mathcal{O}(\alpha_{{\scriptscriptstyle
    \mathrm{S}}})$ calculation of the $y_{{\scriptscriptstyle 12}}$
distribution using approximations for the matrix elements valid in the
soft and quasi-collinear limits.
The Sudakov for the light-quark dipole in
eq.~\eqref{eq:sect22-Delta-qqprime} is, as expected, the same as that
used in ref.~\cite{Hamilton:2012np}, as is the $m_{t}\rightarrow0$
limit of the $bt$ Sudakov form factor in
eq.~(\ref{eq:sect22-Delta-bt-ii}).  It is
easy to see that for $m_{t}^{2}\gg q^{2}$, there is no double-log term
associated to the top quark in $\Delta_{{\scriptscriptstyle bt}}$.
Finally, we note that the component of the massive Sudakov form factor
owing to quasi-collinear radiation from the top-quark is in agreement
with that found in the resummed $k_{t}$-jet rate predictions of
ref.~\cite{Rodrigo:2003ws}.


\subsection{Tuning the \MINLO{} Sudakov form factor}
\label{subsec:Improved-STJ-MiNLO}

Due to the overall Sudakov form factor in the expression for the cross
section, eq.~\eqref{eq:sect22-dsigma-M-in-terms-of-dsigma-NLO-etc},
our \STJ{} predictions do not diverge in those regions of phase space
where the second light parton in the final state (at Born level)
becomes unresolved, but instead exhibit a smooth physical Sudakov
suppression there. Being finite and physical all through phase space
the \STJ{} computation therefore also yields physical predictions for
inclusive \ST{} production, where conventional fixed order \STJ{}
calculations would instead diverge.

In this subsection we state the accuracy of our \STJ{} simulation for
inclusive \ST{} observables, explaining how we have sought
to improve on it, while keeping the NLO accuracy for \STJ{} quantities
intact. We refer to the improved \STJ{} simulation as \STJR{}. Since
the underlying idea at work here is, at some level, rather simple, the
presentation here is kept brief. Expanded explanations of some of the
stated results here can be found in
appendix~\ref{app:Supplementary-details-on-the-MiNLO-procedure}.  The
results of section~\ref{sec:results} can also be somewhat
helpful/illustrative in this respect.

To understand the accuracy of the \STJ{} simulation for \ST{} observables,
we have studied and clarified the correspondence between its cross
section, eq.~\eqref{eq:sect22-dsigma-M-in-terms-of-dsigma-NLO-etc},
and a NLO-matched resummation formula, accurate at next-to-leading
log ($\mathrm{NLL}_{\sigma}$) in the perturbative expansion of the
cross section\footnote{$\mathrm{NLL}_{\sigma}$ resummation includes all terms of the form
  $\frac{1}{y_{12}}\,\bar{\alpha}_{{\scriptscriptstyle \mathrm{S}}}^{n}\,\ln^{m}\frac{Q}{y_{12}}$,
  with $m=2n-1$ and $m=2n-2$.
  $\mathrm{NNLL}_{\sigma}$ resummation further includes all terms with $m=2n-3$.}
(see Apps.~\ref{subsec:Resummed-y12-jet-rate}-\ref{subsec:STJ-MiNLO-predictions-for-ST-observables}).
We determine that the source of differences between the two starts,
expectedly, at the level
of $\mathrm{NNLL}_{\sigma}$ terms. On integration over $y_{12}$
these $\mathrm{NNLL}_{\sigma}$ differences give rise to a distribution
of Born kinematics different to that of conventional NLO \ST{} by
terms of order
\begin{equation}
\int_{0}^{Q^{2}}dy_{12}\,\Delta(y_{12})\,\frac{d\sigma_{{\scriptscriptstyle \mathrm{LO}}}^{{\scriptscriptstyle \mathrm{ST}}}}{d\Phi}\,\bar{\alpha}_{{\scriptscriptstyle \mathrm{S}}}^{n}\,\frac{1}{y_{12}}\ln^{m}\frac{Q^{2}}{y_{12}}\,=\,\frac{d\sigma_{{\scriptscriptstyle \mathrm{LO}}}^{{\scriptscriptstyle \mathrm{ST}}}}{d\Phi}\,\cdot\,\mathcal{O}(\bar{\alpha}_{{\scriptscriptstyle \mathrm{S}}}^{n-\frac{m+1}{2}})\,,\label{eq:sect23-Qbt-eq-two-pb-dot-pt}
\end{equation}
with $n\ge2$ and $m=2n-3$ in the case of $\mathrm{NNLL}_{\sigma}$ terms. In other
words, the \MINLO-improved \STJ{} simulation has only LO accuracy for
\ST{} observables, a fact well supported by our numerical studies in
section~\ref{sec:results}.

Given that the \STJ{} formula,
eq.~\eqref{eq:sect22-dsigma-M-in-terms-of-dsigma-NLO-etc},
already contains, through factorization at the one-loop level, the
process-dependent virtual corrections to \ST{} production, we postulate
that the only modification needed to promote it to $\mathrm{NNLL}_{\sigma}$
accuracy is the extension of the Sudakov form factor to that
order.\footnote{As has been the case
  for all \MINLO{} simulations that have been proven to reach NLO accuracy
  for the associated lower multiplicity process so far~\cite{Hamilton:2012rf,Luisoni:2013kna,Frederix:2015fyz,Hamilton:2016bfu}}
If such an extension were then to be implemented in the \STJ{} simulation
its $y_{12}$ distribution would converge on that of the $\mathrm{NNLL}_{\sigma}$
resummation, at the same time eliminating those terms which caused
the distribution of its inclusive Born kinematics to deviate from
NLO by a relative $\mathcal{O}(\bar{\alpha}_{{\scriptscriptstyle \mathrm{S}}})$
amount (eq.~\eqref{eq:sect23-Qbt-eq-two-pb-dot-pt}). Residual
$\mathrm{N^{3}LL}_{\sigma}$ differences ($m=2n-4$) will instead mean
that the latter deviations reduce to relative
$\mathcal{O}(\bar{\alpha}_{{\scriptscriptstyle \mathrm{S}}}^{3/2})$.
This point is elaborated on in appendix~\ref{subsec:Improving-STJ-MiNLO}.

We note that it is possible, in principle, to adjust the coefficient
of the $\bar{\alpha}_{{\scriptscriptstyle \mathrm{S}}}^2 \ln^2(Q^{2}/y_{12})$ term in the Sudakov form factor
by a formally subleading $y_{12}$-independent factor,
$\sim1+\mathcal{O}(\sqrt{\bar{\alpha}_{{\scriptscriptstyle
      \mathrm{S}}}})$, such that the distribution of the \ST{} Born
kinematics returned by the \STJ{} calculation ($\Phi$) becomes
identical to $d\sigma_{{\scriptscriptstyle
    \mathrm{NLO}}}^{{\scriptscriptstyle \mathrm{ST}}}/d\Phi$.  We
further note that the latter form of the suggested
$\mathrm{NNLL}_{\sigma}$ Sudakov form factor extension is precisely
what one would obtain by fitting the $\mathcal{O}(1)$ function
$\mathcal{A}_{2}(\Phi)$ inside
\begin{equation}
\ln\delta\Delta(y_{12})\,=\,-2\int_{y_{12}}^{Q_{bt}^{2}}\frac{dq^{2}}{q^{2}}\,\bar{\alpha}_{{\scriptscriptstyle \mathrm{S}}}^{2}\,\mathcal{A}_{2}(\Phi)\,\ln\frac{Q_{bt}^{2}}{q^{2}}\,,\label{eq:sect23-log-delta-Delta-sudakov-correction-factor}
\end{equation}
such that 
\begin{equation}
\frac{d\sigma_{{\scriptscriptstyle \mathrm{NLO}}}^{{\scriptscriptstyle \mathrm{ST}}}}{d\Phi}\,=\,\int dy_{12}\,\frac{d\sigma_{{\scriptscriptstyle \mathcal{M}}}}{d\Phi dy_{12}}\,\delta\Delta(y_{12})\,.\label{eq:sect23-dsigNLO-eq-int-dsigMiNLO-times-delta-Delta}
\end{equation}
We have chosen to normalise $\ln\delta\Delta(y_{12})$ with the factor
of two on the right-hand side of
eq.~\eqref{eq:sect23-log-delta-Delta-sudakov-correction-factor}, to
account for the fact that $t$-channel single-top production consists
of two emitting quark dipoles at lowest order, to enable a more
easy/meaningful comparison with other typical Sudakov coefficients
at the same order.

Eq.~\eqref{eq:sect23-dsigNLO-eq-int-dsigMiNLO-times-delta-Delta}
summarises the improvement procedure which we have applied to our
baseline \STJ{} construction described in
sections~\ref{subsec:STJ-NLO}-\ref{subsec:STJ-MiNLO}.
The fit procedure to arrive at $\mathcal{A}_{2}(\Phi)$ in
eq.~\eqref{eq:sect23-log-delta-Delta-sudakov-correction-factor}
can be attempted in a variety of ways, and we have chosen to use
an advanced procedure based on neural network techniques, for which
we give details in section~\ref{sec:NN}. In practice, we have
implemented the Sudakov form factor correction, $\delta\Delta(y_{12})$,
evaluating the $q^{2}$ integral in
eq.~\eqref{eq:sect23-log-delta-Delta-sudakov-correction-factor}
with a one-loop running coupling:
\begin{equation}
\qquad\ln\delta\Delta(y_{12})\,=\,\mathcal{A}_{2}(\Phi)\,\mathcal{G}_{2}(\lambda)\,,\qquad\qquad\mathcal{G}_{2}(\lambda)\,=\,\frac{-1}{2\pi^{2}\beta_{0}^{2}}\,\left[\frac{2\lambda+(1-2\lambda)\ln(1-2\lambda)}{1-2\lambda}\right]\,,
\label{eq:sect23-lndeltaDelta-in-terms-of-lambdas}
\end{equation}
with 
\begin{equation}
\beta_{0}\,=\,\frac{11C_{A}-2n_{f}}{12\pi}\,,\quad\qquad\lambda=\frac{1}{2}\alpha_{{\scriptscriptstyle \mathrm{S}}}\beta_{0}\ln\frac{Q_{bt}^{2}}{y_{12}}\,.
\label{eq:sect23-beta0-defn-lambda-defn}
\end{equation}

While we use a form for the $y_{12}$ resummation
formula at $\mathrm{NNLL}_{\sigma}$, we do not presume to know the
details of the related Sudakov ingredients at that order, so we assume
that the $\mathcal{A}_{2}$ coefficient has a general dependence on
$\Phi$ already for this reason. There are, however, established grounds
to expect $\mathcal{A}_{2}$ to be generically $\Phi$-dependent,
as also elaborated in appendix~\ref{subsec:Improving-STJ-MiNLO}.

We also point out that, if it is the case that the differences between
the \STJ{} and NLO \ST{} $\Phi$-distributions owe purely to the
omission of terms in the Sudakov form factor, the $\mathcal{A}_{2}$
function fitted in this work, for a given 13 TeV LHC setup, should
remain valid for different beam energies, PDF sets, etc. We have
carried out empirical investigations regarding this point, using the
\STJR{} \MINLO{} Sudakov form factor fitted using samples of \ST{}
and \STJ{} generated for a 13 TeV LHC, to make predictions at 8 TeV.
We find that the \STJR{} simulation with the latter fit reproduces
inclusive 8 TeV $t$-channel single-top observables remarkably well.
A representative sample of results from that study is given in
appendix~\ref{app:Tune-extrapolation}. Indeed, the latter results
strongly suggest that a dedicated refitting of the \STJR{} Sudakov
form factor, using 8 TeV \ST{} and \STJ{} events, would fare
comparably to the one based on fitting with 13 TeV events.

Should the missing $\mathrm{NNLL}_{\sigma}$ terms in the \MINLO{}
Sudakov form factor of \STJ{} not fully account for the leading
($\mathcal{O}(\bar{\alpha}_{{\scriptscriptstyle \mathrm{S}}})$)
deviations in its $\Phi$ distribution with respect to NLO \ST{}
predictions, the modification in
eq.~\eqref{eq:sect23-dsigNLO-eq-int-dsigMiNLO-times-delta-Delta}
is still admissible, provided that the fitted $\mathcal{A}_{2}$ is
of the same order of magnitude as other Sudakov coefficients. It
does not compromise the NLO accuracy of the \STJ{} generator for
$t$-channel single-top plus jet observables, or change its
resummation accuracy.

Finally, it is reasonable to ask why we have chosen to use $Q_{bt}$
for the hard scale in
eq.~\eqref{eq:sect23-log-delta-Delta-sudakov-correction-factor} rather
than $Q_{qq^{\prime}}$. Without a much more sophisticated NLO
calculational framework, wherein one has the ability to clearly
distinguish which contributions to the NLO cross section are
associated to which colour dipole in the leading order process
($qq^{\prime}/bt$), it is not possible to carry out the correction
procedure proposed here on a dipole-by-dipole basis. Hence, we are
limited to having one cross section unitarity constraint which we can
use to fit one term in Sudakov form factor. This does not pose a great
problem in practice, since regions of the \ST{} Born phase space where
$Q_{bt}$ can be disparate from $Q_{qq^{\prime}}$ are strongly
suppressed. Moreover, in the context of the \STJ{} generator, the
great bulk of events populating such regions are always anyhow subject
to large Sudakov logarithms associated with soft corrections to the
$bt$ system. We have assessed ambiguities related to choosing $Q_{bt}$
as the hard scale in $\delta\Delta(y_{12})$ conservatively (varying
$Q_{tb}$ up and down by a factor of four), finding a negligible
impact in all of the $\mathcal{O}(200)$ distributions considered in
our studies. These uncertainties are depicted in \textit{all} of the
plots of our results section, \ref{sec:results}, as dark red bands,
but for the most part they are so small as to be invisible.

\subsection{Neural network fit}\label{sec:NN}

In this section we describe how we have fitted the
$\mathcal{A}_{2}(\Phi)$ Sudakov coefficient in
eqs.~(\ref{eq:sect23-log-delta-Delta-sudakov-correction-factor}-%
\ref{eq:sect23-lndeltaDelta-in-terms-of-lambdas}), through
imposing the differential unitarity constraint expressed in
eq.~\eqref{eq:sect23-dsigNLO-eq-int-dsigMiNLO-times-delta-Delta}.
With the fitted $\mathcal{A}_{2}(\Phi)$ in hand we then simply
reweight \STJ{}$\rightarrow$\STJR{} events by multiplying
them with the Sudakov form factor correction, $\delta\Delta(y_{12})$,
as in the integrand on the right-hand side of 
eq.~\eqref{eq:sect23-dsigNLO-eq-int-dsigMiNLO-times-delta-Delta}.

To quicken the development of the method and give it much greater
flexibility, in implementing our tuning procedure for the \MINLO{}
Sudakov form factor, we have chosen to define the Born variables,
$\Phi$, slightly differently to how they were introduced at the
beginning of section~\ref{subsec:STJ-MiNLO}. For the purposes
of this part of the work, they are defined from the set of momenta
that result from applying the exclusive $k_{t}$ algorithm to the
events in the Les Houches files output by the \ST{} and \STJ{}
generators; rather than from directly clustering the underlying
Born configurations in the case of the \STJ{} simulation. It is
natural to expect that differences resulting from this modification
are small, since the effective clustering represented by the inverse
of the \POWHEGBOX{} \FKS{} mapping~\cite{Frixione:1995ms,Frixione:2007vw}
is based on the transverse momentum separation of partons in the soft
and collinear limits, as in the $k_t$ algorithm. This is nevertheless
an approximation, made of convenience rather than necessity. However,
as noted at the end of section~\ref{subsec:Improved-STJ-MiNLO}, and
as we shall go on to demonstrate in section~\ref{sec:results}, the
\MINLO{} tuning procedure we have carried out is remarkably robust
even against strong variations in its associated parameters, e.g.
the hard scale in $\delta\Delta(y_{12})$,
eq.~\eqref{eq:sect23-log-delta-Delta-sudakov-correction-factor}.

Being a $2\rightarrow2$ scattering process at leading order, the
Born phase space of $t$-channel single-top production can be
parametrized using three independent variables. There is freedom
in the selection of these variables, and we have opted for simple
quantities which are rather uncorrelated from one another. As stated
above, the Born variables are defined from the two momenta associated
with the final-state after clustering the events in the \ST{} and
\STJ{} Les Houches files with the exclusive $k_t$ algorithm,
until each one consists of just the top quark and a light jet, whose
momenta we label $p_t$ and $p_j$.\footnote{
  Rarely an event will fail to cluster back to a two-body \ST{} final-state,
  due to the flavour conservation implemented in our $k_t$ clustering
  (footnote \ref{footnote:flavour-sensitive-kt-clustering}). 
  Having no associated \ST{} Born configuration, such events are omitted
  from the ${\mathcal{A}}_{2}(\Phi)$ fitting procedure, and are untouched
  by the related reweighting.
}
From the latter we construct our chosen Born variables: $y_{tj}$,
the rapidity of $p_{t}+p_{j}$; $\hat y_{t}$, the rapidity interval
between the top quark and the latter; and $p_{t,\mathrm{top}}$, the
transverse momentum of the top.

Since the constraint to be solved for $\mathcal{A}_{2}(\Phi)$,
eq.~\eqref{eq:sect23-dsigNLO-eq-int-dsigMiNLO-times-delta-Delta},
involves a \emph{convolution} of the \STJ{} cross section with
$\delta\Delta(y_{12})$,
eq.~\eqref{eq:sect23-lndeltaDelta-in-terms-of-lambdas}, we discretize
the three dimensional space spanned by the Born variables, in order to
use eq.~\eqref{eq:sect23-dsigNLO-eq-int-dsigMiNLO-times-delta-Delta}
to fit $\mathcal{A}_{2}(\Phi)$ using samples of \ST{} and \STJ{} Les
Houches events.\footnote{An alternative reweighting method, based on weighted kernel
  density estimation, has been explored as well. In this method the
  discretization of the phase-space can be avoided, but it comes
  with a huge computational cost and large memory usage and has
  therefore been disregared.}
 The discretization is carried out by first creating a
regular binning in the physically accessible region of the $\hat y_{t}
- y_{tj}$ plane, wherein each bin covers $0.5 \times 0.5$ units of
rapidity. Each of the latter 2D bins is further segmented
according to $p_{t,\mathrm{top}}$, in such a way that all resulting
bins in the three dimensional parameter space contain 2000 of the 18
million \STJ{} events used in carrying out the subsequent fit.%
\footnote{At the edges of the phase space adjacent bins are
  combined, iteratively, if they are found to contain less than
  2000 events.} %

The fitted $\mathcal{A}_{2}(\Phi)$ function is then determined by
minimizing the following loss function:
\begin{equation}
  \mathcal{L} \,\, = \,\, \sum_{i=1}^{N_{\rm bins}} \, %
  \left[\, %
  \sum_{j=1}^{N} \, w^{\scriptscriptstyle {\rm ST}}_{i,j} %
  \, - \,%
  \sum_{k=1}^{N'} \, %
                   w^{\scriptscriptstyle{\rm STJ}}_{i,k} \,%
                   \mathrm{e}^{\,%
                     \tilde{\mathcal{A}_{2}}(\Phi_{i})%
                     \,%
                     \mathcal{G}_{2}(\lambda) %
                              }%
  \,\right]^2.\label{eq:sect25-loss-fn-defn}
\end{equation}
Here, in eq.~\eqref{eq:sect25-loss-fn-defn}, $N_{\rm bins}$ is the
total number of bins in the discretized three dimensional Born
variable parameter space. $N$ and $N'$ are, respectively, the
number of \ST{} and \STJ{} events used in carrying out the fit.
$w^{\scriptscriptstyle{\rm ST}}_{i,j}$ is the weight of the $j$th
\ST{} event in bin $i$ of the discretized Born variable parameter
space, with $w^{\scriptscriptstyle{\rm STJ}}_{i,k}$ being analogously
defined for the \STJ{} events. $\tilde{\mathcal{A}_{2}}(\Phi)$ in
eq.~\eqref{eq:sect25-loss-fn-defn} is the model prediction for
the desired effective Sudakov form factor coefficient
(eqs.~(\ref{eq:sect23-log-delta-Delta-sudakov-correction-factor}-%
\ref{eq:sect23-lndeltaDelta-in-terms-of-lambdas})), evaluated
at the centre of bin $i$. $\mathcal{G}_{2}(\lambda)$ is as defined
in eq.~\eqref{eq:sect23-lndeltaDelta-in-terms-of-lambdas}.

The fit of $\mathcal{A}_{2}(\Phi)$ according to
eq.~\eqref{eq:sect25-loss-fn-defn} is performed with machine
learning techniques. To avoid making assumptions regarding the
analytic form of $\mathcal{A}_{2}(\Phi)$, we have employed an
artificial neural network parametrization based on a feed-forward
multi-layer perceptron.
This choice eliminates the requirement of selecting a specific
functional form for our problem, by providing a non-linear model
which learns the data structure. Additionally, the
$\mathcal{A}_{2}(\Phi)$ function fitted in this way evaluates
quickly when called on to reweight \STJ{}$\rightarrow$\STJR{}
events with the Sudakov form factor correction, $\delta\Delta(y_{12})$.

A grid search was carried out to determine the best neural network
architecture, loss function definition, and optimizer algorithm for
our framework. The best architecture was found to consist of a neural
network with two hidden layers, comprising five and three nodes
respectively, based on hyperbolic tangent activation
functions.~\footnote{In the grid search procedure several models and
  training setups are tested and the setup which obtains the lowest
  cost function is proposed as the best model.} The output layer for
the architecture consists of a single node with a linear activation
function.  In total this model requires the tune of 42 parameters in
the form of weights and biases. An efficient genetic optimizer was
implemented to train the model, based on the covariance matrix
evolution strategy (CMA-ES) \cite{Hansen:2001ecj}.

Fits to the $\mathcal{A}_{2}(\Phi)$ function were carried
out with the latter neural network setup, using samples
of 25 million \ST{} and 18 million \STJ{} \POWHEG{} Les
Houches events, at a 13 TeV LHC.\footnote{Details on the
  values of physical constants and other parameters used
  to generate these samples follow at the beginning of
  section~\ref{sec:results}.}
The same setup and statistics were also used to perform
analogous fits for the case of single-anti-top production.
Both for $t$-channel single-top and single-anti-top processes
eleven $\mathcal{A}_{2}(\Phi)$ fits were carried out. Seven of
these correspond to redoing the fit in the presence of correlated
renormalization and factorization scale variations in the \ST{}
and \STJ{} generators; i.e.~we perform a separate fit of
$\mathcal{A}_{2}(\Phi)$ varying
$\mu_{\scriptscriptstyle R/F} \rightarrow K_{\scriptscriptstyle R/F} \,%
\mu_{\scriptscriptstyle R/F}$,
in both the \ST{} and \STJ{} simulations, for all pairings of
$\KR$ and $\KF$ values in $\{\frac{1}{2},\,1,\,2\}$, discarding
the two pairings where $\KR$ and $\KF$ differ by more than a
factor of two. For the central scale choice, four further $\mathcal{A}_{2}(\Phi)$ fits are
carried out for which the scale $Q_{bt}$ in $\delta\Delta(y_{12})$,
eq.~\eqref{eq:sect23-log-delta-Delta-sudakov-correction-factor},
is multiplied by
$K_{\scriptscriptstyle{Q_{bt}}}\,\in\,\{\frac{1}{4},\,\frac{1}{2},\,2,\,4\}$,
in order to gauge sensitivity to that scale choice.

\begin{figure}
\begin{center}
  \includegraphics[width=0.327\textwidth]{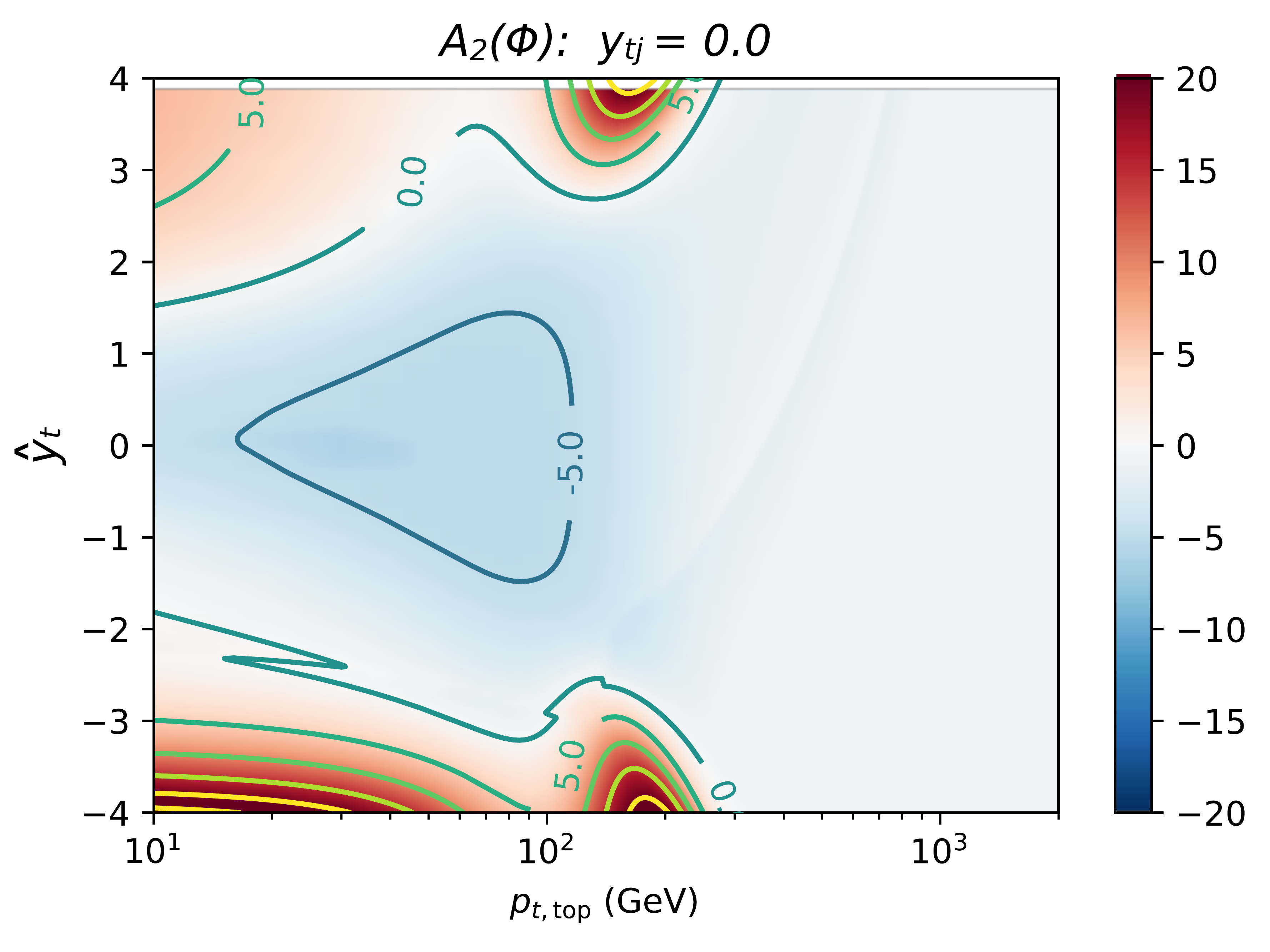}
  \includegraphics[width=0.327\textwidth]{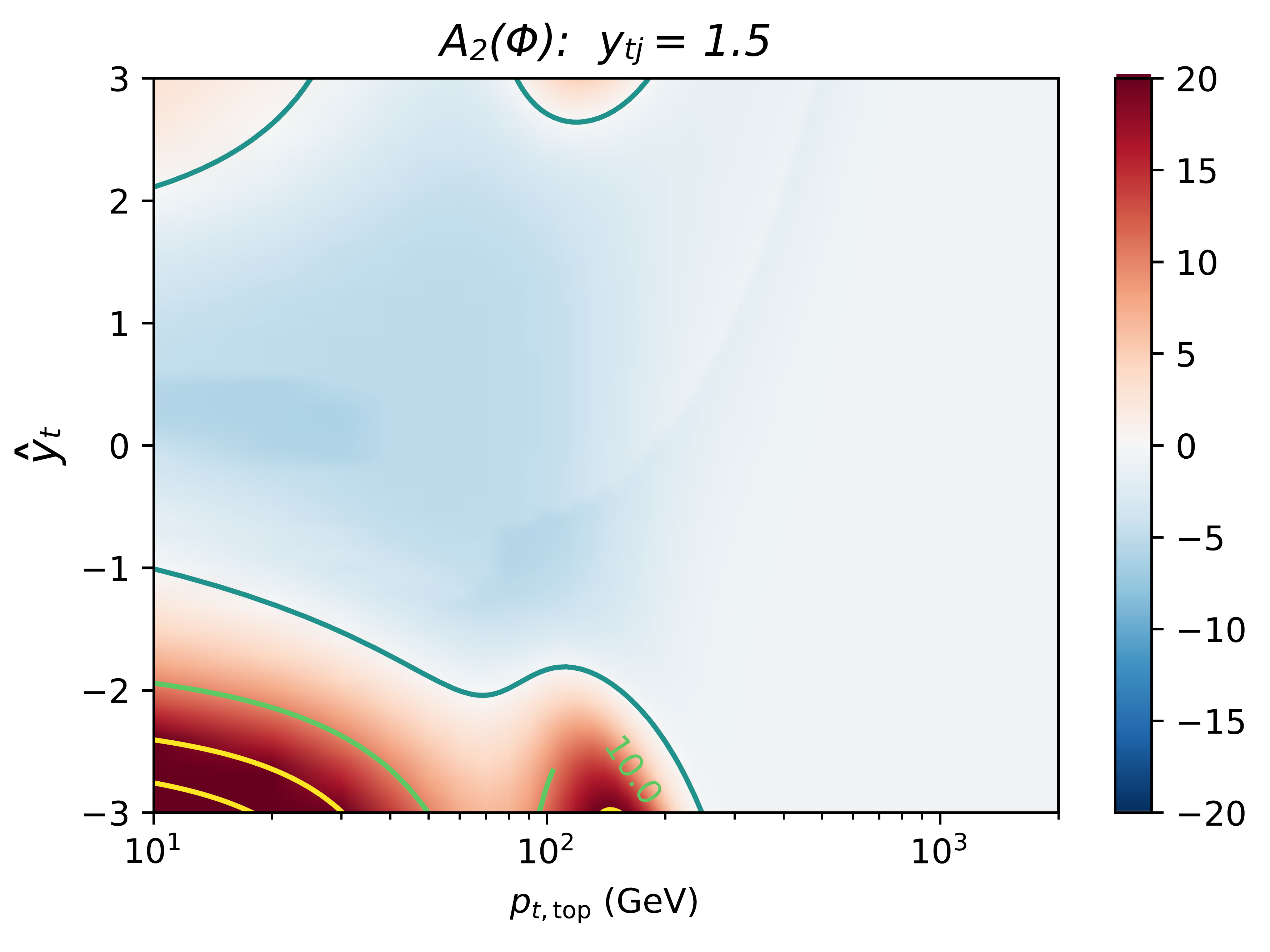}
  \includegraphics[width=0.327\textwidth]{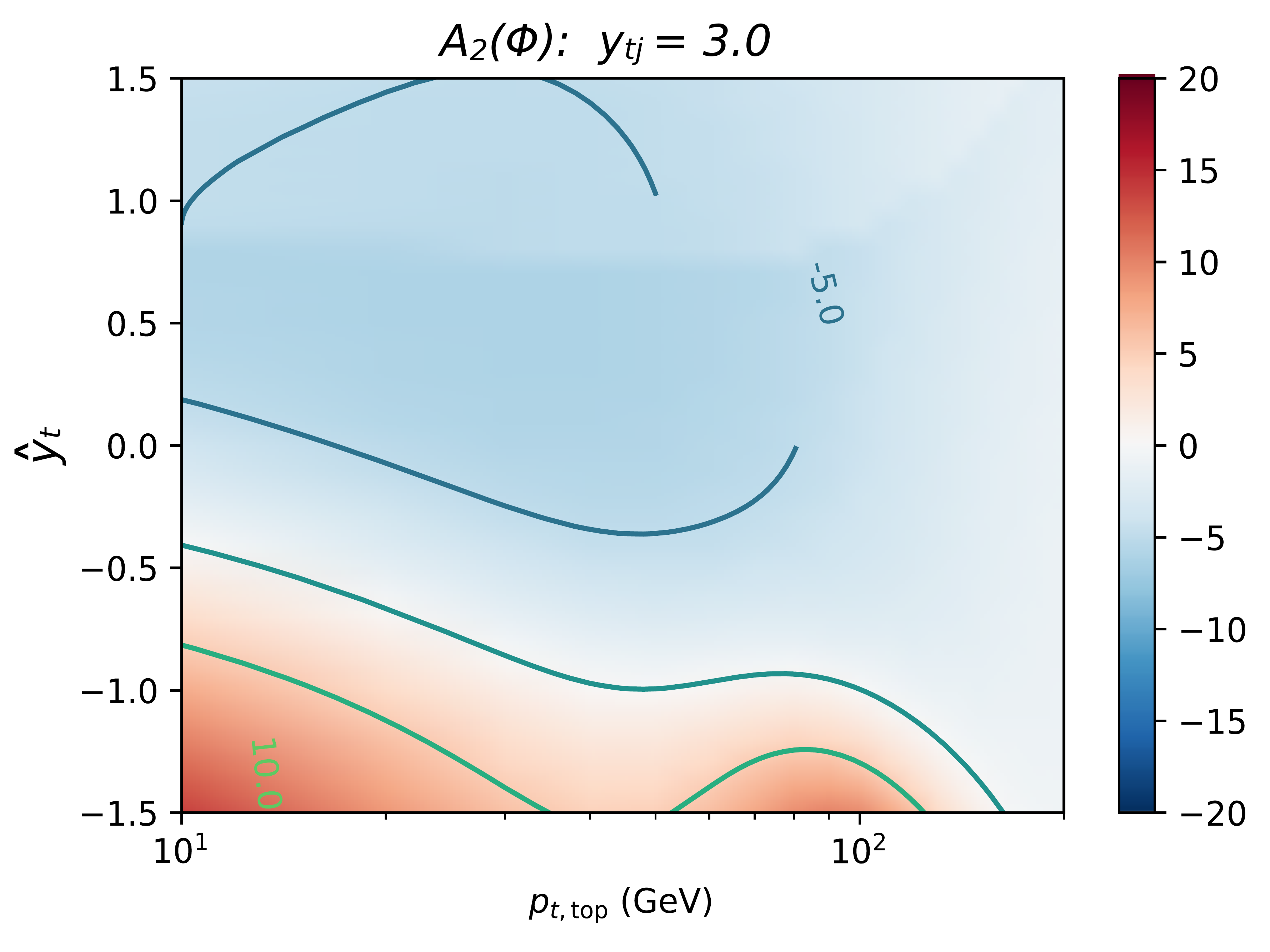}
  \caption{Heatmap plots of the fitted $\mathcal{A}_{2}(\Phi)$
    term in the \MINLO{} Sudakov form factor, defined through
    eqs.~(%
    \ref{eq:sect23-log-delta-Delta-sudakov-correction-factor}%
    -%
    \ref{eq:sect23-lndeltaDelta-in-terms-of-lambdas}); %
    used in promoting \STJ{}$\rightarrow$\STJR{} events
    by reweighting the former with the $\mathrm{NNLL}_\sigma$
    exponential factor, $\delta\Delta(y_{12})$
    (eq.~\eqref{eq:sect23-lndeltaDelta-in-terms-of-lambdas}).
    The $\mathcal{A}_{2}(\Phi)$ shown here has been obtained
    using the default (central) scale choices in the \ST{}
    and \STJ{} generators, for $y_{tj}=0.0$ (left plot),
    $y_{tj}=1.5$ (centre plot) and $y_{tj}=3.0$ (right plot).
    The ranges in each of the three plots vary in order
    to limit the amount of physically inaccessible phase
    space shown, while not cutting away any accessible regions.
  }
  \label{NNOutput}
\end{center}
\end{figure}

In Fig.~\ref{NNOutput}
we project the trained neural network model obtained with the setup described in section~\ref{sec:setup}, for the central
renormalization and factorization scale choice in $t$-channel
single-top production, into the $\hat{y}_{t}-\,p_{t,\rm top}$
plane, for $y_{tj}=0.0$ (left plot), $y_{tj}=1.5$ (centre plot),
and $y_{tj}=3.0$ (right plot). To gain some perspective on the
size of the $\mathcal{A}_{2}$ values shown in the heatmap plots
of Fig.~\ref{NNOutput}, we point out that the function
multiplying it in the Sudakov form factor in our work,
$\mathcal{G}_{2}(\lambda)$,
eq.~\eqref{eq:sect23-lndeltaDelta-in-terms-of-lambdas}, is
precisely  the same as that multiplying the $A_2$ coefficient
in eq.~(10b) of ref.~\cite{Banfi:2012jm}; modulo an extra factor
of two in our case, accounting for the fact that we have two
colour dipoles in our lowest order process, while those considered
in ref.~\cite{Banfi:2012jm} consist of just one. In
ref.~\cite{Banfi:2012jm} $A_{2} \simeq 9$ for the Drell-Yan
process, and $A_{2} \simeq 21$ in the case of Higgs production
via gluon fusion.
We conclude that the fitted $\mathcal{A}_{2}(\Phi)$ is numerically of
similar size to these $A_{2}$ coefficients in the entire phase space.

\section{Results}\label{sec:results}

In the following we discuss a representative sample of distributions
obtained in the context of our validation of the new \STJ{} \NLOPS{}
generator, as set out in
sections~\ref{subsec:definition}\,-\,\ref{subsec:STJ-MiNLO}, along with
its improved \STJR{} counterpart, based on the fitting of the Sudakov
form factor described in sections~\ref{subsec:Improved-STJ-MiNLO}\,-\,%
\ref{sec:NN}. We remind that both \STJ{} and \STJR{} simulations aim
at NLO accuracy in the description of $t$-channel single-top plus
jet events, while the latter is also intended to be NLO accurate in
the description of generic inclusive $t$-channel single-top observables.
The \POWHEG{} \NLOPS{} simulation of $t$-channel single-top production,
\ST{} \cite{Alioli:2009je}, is used throughout to assess the quality
of the description afforded by \STJ{} and \STJR{} for inclusive
quantities, and to gauge the magnitude of NLO effects in $t$-channel
single-top plus jet events.

\subsection{Setup}\label{sec:setup}
Both here in our validation studies and in fitting the \MINLO{}
Sudakov form factor, we have considered 13 TeV LHC collisions.
We use the NNLO \NNPDF{}~3.0 parton distribution
functions~\cite{Ball:2014uwa} corresponding to $\alphaS(m_{Z}) = 0.118$ via the \LHAPDF{}
package~\cite{Buckley:2014ana} (index 261000).
%
The Fermi constant is set to $G_F = 1.16639 \times 10^{-5}$
GeV$^{-2}$. The $Z$-boson mass is set to $m_{Z} = 91.118$~GeV,
and the fine structure constant evaluated at that scale is
given by $1/\alpha = 127.012$. The $W$-boson mass, the
weak mixing angle, and the weak coupling constant are hence
derived according to tree-level relations among the
electroweak parameters.
%
The top quark mass has been set to $172.5$ GeV, while all
other quark masses have been set to zero.

We use a diagonal CKM matrix. On the heavy quark line, where
the bottom quark converts to a top quark, we therefore have
$V_{tb} = 1$, which is well within the uncertainties on
its current determination from Tevatron and LHC data
$|V_{tb}| = 1.009 \pm 0.031$~\cite{Patrignani:2016xqp}.
If we further sum over the flavours of the final-state quark on the
associated light quark line, since the CKM matrix is unitarity, our matrix
elements will be identical to those obtained with the full CKM
matrix.

All results shown in this section include the effects of
parton showering, simulated with the
\PYTHIAEIGHT{} program~\cite{Sjostrand:2014zea}. Since our
primary intention is to validate the new \STJ{} generator
and its tuned \STJR{} counterpart, we have switched off
hadronization and multiple parton interactions in \PYTHIAEIGHT{},
and we treat the top quark as a stable particle.%
\footnote{
  We remind that one can relatively quickly produce new
  Les Houches event files in which the top quarks have
  been decayed, according to the relevant tree-level matrix
  elements, using \MadSpin{}~\cite{Frixione:2007zp,Artoisenet:2012st}.
}
We have found it important to adopt a new momentum reshuffling
option in \PYTHIAEIGHT{},%
\footnote{We set {\texttt{SpaceShower:dipoleRecoil = on}} in the
  \PYTHIAEIGHT{} input file.  Ref.~\cite{Cabouat:2017rzi} describes in
  detail the physical reasoning behind this option and how it modifies
  the showering of initial-final QCD dipoles. It states that this option
  is theoretically better motivated than its alternatives.} intended
to yield an alternative treatment of showering initial-final QCD
dipoles~\cite{Cabouat:2017rzi}. Similar findings in recent studies on
vector boson scattering simulations have been commented on in
ref.~\cite{Ballestrero:2018anz}.

In the \ST{} simulation the central renormalization and
factorization scale choice is $\muR = \muF = m_{t}$.
Theoretical uncertainties are estimated by varying $\muR$
and $\muF$, independently, up and down by a factor of two,
while keeping $\frac{1}{2} \le \muR/\muF \le 2$. The
envelope of the predictions following from these variations
defines the theoretical uncertainty. For \STJ{} the central
scale choice is dictated by the \MINLO{} prescription in
section~\ref{subsec:STJ-MiNLO}, with uncertainties being
estimated in complete analogy to the \ST{} case. The theoretical
uncertainties for \STJR{} follow as in the \STJ{} case, but
with $\mathcal{A}_{2}(\Phi)$ changing according to $\muR$
and $\muF$, to maintain the equality in
eq.~\eqref{eq:sect23-dsigNLO-eq-int-dsigMiNLO-times-delta-Delta},
with $\muR$ and $\muF$ being varied about their central values
in the same way on both sides of that equation.

As mentioned at the end of sections~\ref{subsec:Improved-STJ-MiNLO}
and \ref{sec:NN}, we also investigate the uncertainty in \STJR{}
predictions owing to the ambiguity in choosing $Q_{bt}$ as the
hard scale in the $\mathrm{NNLL}_{\sigma}$ reweighting factor
$\delta\Delta(y_{12})$
(eq.~\eqref{eq:sect23-log-delta-Delta-sudakov-correction-factor}).
This uncertainty is estimated by taking the envelope of predictions
obtained by rescaling $Q_{bt}$ in $\delta\Delta(y_{12})$ up and
down by a factor of four, fitting a new $\mathcal{A}_{2}(\Phi)$
for each $Q_{bt}$ variation, so as to maintain
eq.~\eqref{eq:sect23-dsigNLO-eq-int-dsigMiNLO-times-delta-Delta}.
This uncertainty is almost always too small to be visible in
our results, and never exceeds that due to renormalization
and factorization scale variation.

Finally, in validating our simulations, we have studied the same
extensive range of distributions obtained from an 8 TeV LHC setup,
identical to the 13 TeV one described above. The 8 TeV analysis
was carried out without refitting the \MINLO{} Sudakov form factor
for \STJR{}, which remains the same as in the 13 TeV study
immediately following below. A representative subset of these 8 TeV
predictions is deferred to appendix~\ref{app:Tune-extrapolation},
to avoid repetition, since the findings are very much the same as
for the 13 TeV case discussed here, in
sections~\ref{subsec:top-rapidity-and-pT}-%
\ref{subsec:top-transverse-momentum-in-single-jet-events}.

\subsection{Guide to plots}

All plots in this section show predictions, with uncertainty
estimates, from the \ST{} simulation in green, \STJ{} in blue, and
the tuned \STJR{} generator in red. In each case the top panel shows
absolute cross section predictions, while the lower three panels
display ratios of the various results to one another.
$Q_{bt}$ variation in $\delta\Delta(y_{12})$
(eq.~\eqref{eq:sect23-log-delta-Delta-sudakov-correction-factor})
is depicted by dark red shading, but is often too small to be visible. 

The order of the presentation roughly follows the degree of
exclusivity of the studied observables, starting with the most
inclusive, for which the \ST{} and \STJR{} simulations should
be NLO accurate, working towards more exclusive quantities,
for which \STJ{} and \STJR{} should provide the best predictions.

\subsection{Top quark rapidity and transverse momentum}
\label{subsec:top-rapidity-and-pT}

\begin{figure}
\begin{center}  
  \includegraphics[width=0.459\textwidth]{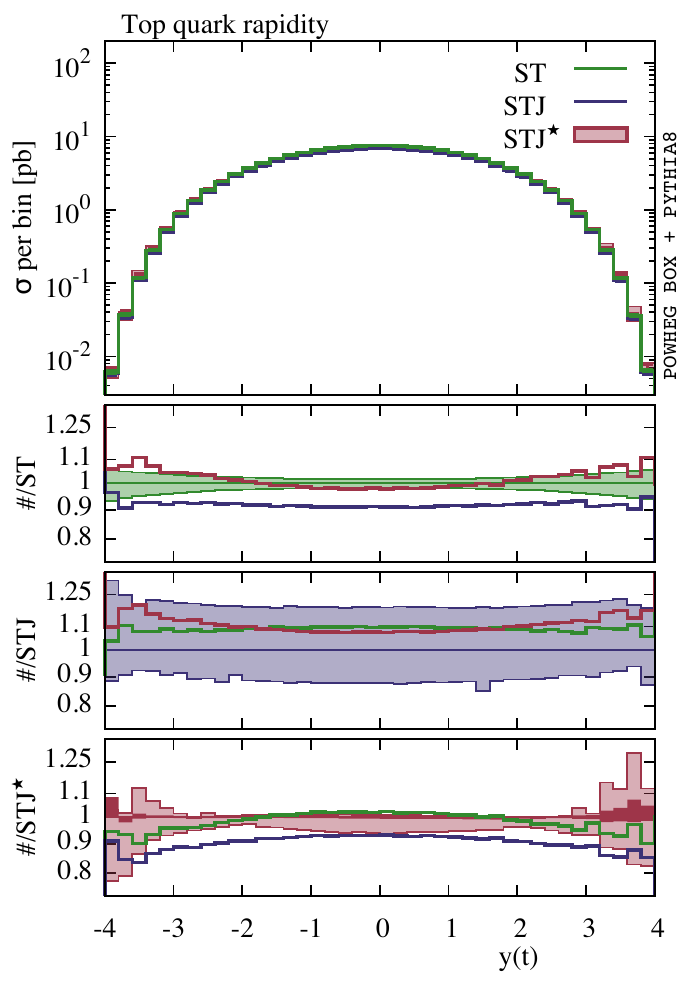}  
  \hspace{10 mm}
  \includegraphics[width=0.459\textwidth]{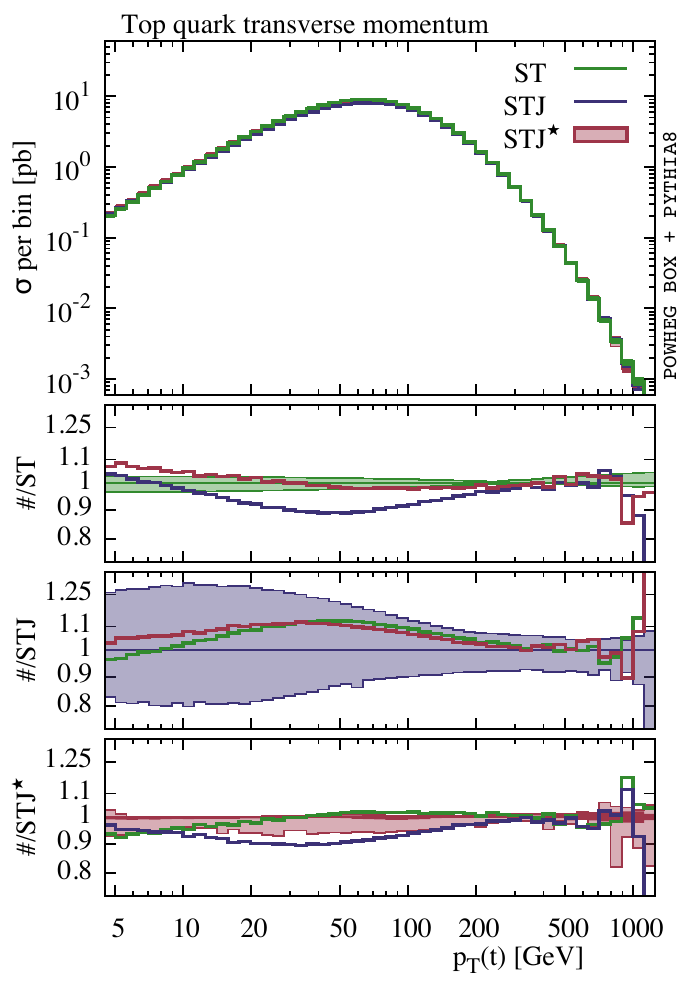}
\end{center}
\caption{Rapidity (left) and transverse momentum (right) of
  the top quark in $t$-channel single-top production. Predictions
  from the \POWHEG{} \ST{} program~\cite{Alioli:2009je} are shown
  in green. Results from the new \MINLO{} \STJ{} simulation
  are displayed in blue, while those of its improved counterpart,
  \STJR{}, appear in red. All predictions include parton shower
  effects simulated by \PYTHIAEIGHT{}~\cite{Sjostrand:2014zea}.
} 
\label{fig:top}
\end{figure}

In Fig.~\ref{fig:top} we present predictions for the
top quark rapidity and transverse momentum distributions.
Being inclusive with respect to all jet activity, the \ST{} simulation
(green) provides NLO accurate predictions for these observables, while
\STJ{} (blue) is formally only LO accurate
(section~\ref{subsec:Improved-STJ-MiNLO} and
appendix~\ref{subsec:STJ-MiNLO-predictions-for-ST-observables}).  This
statement is substantiated by the two distributions in
Fig.~\ref{fig:top}. The inclusive
\ST{} predictions carry a remarkably small QCD scale uncertainty at
NLO, as is well known to be the case for inclusive $t$-channel
single-top observables, with the \STJ{} predictions lying no more than
10\% away from the latter, throughout almost all of the two
distributions, and with a larger associated uncertainty, compatible
with the fact that it is only LO accurate.

In the case of the top quark rapidity distribution the improved
\STJR{} simulation agrees with the \ST{} results to within
$\lesssim2\%$ in the central region, deviating slightly from it,
by $\sim6\%$, at high values of the absolute rapidity, $|y(t)|>3$.
These deviations are, nevertheless, just of the same size as the \ST{}
scale uncertainties in these regions, modulo some statistical
fluctuations.

Besides the central prediction of \STJR{} converging on that of
the \ST{} simulation, so too does its scale uncertainty band. The
uncertainty band of the \ST{} simulation is as small as $\pm3\%$
in the central $y(t)$ region of the first ratio plot. The \STJR{}
uncertainty band in the third ratio plot is at the level of
$+2\%$/$-6\%$ in the same region, to be compared with $+20\%$/$-10\%$
in the \STJ{} case.

At the extremities of the top quark rapidity distribution,
$|y(t)|\gtrsim3.5$, the \STJR{} uncertainty band exceeds that of
\ST{}, and looks somewhat more like that of \STJ{}. Such imperfections
are not entirely unexpected in these regions due to limited
statistics, especially when working with the weighted events that
determine the scale variations, which carry greater statistical noise
in the neural network fitting procedure than those determining the
central prediction. The discretization of the Born variable parameter
space used in the fit can also become coarse in these lowly populated
high-rapidity regions. In addition, it is worth remembering that the
neural network model makes no assumptions, whatsoever, on the form of
the function to be fitted, and is ultimately limited to just 42
parameters.

Further refinement and/or complexity in our \STJR{} neural
network model, could increase the level of convergence of the
\STJR{} uncertainty band to that of \ST{}. However, the
improvement in both the description of the central value and
the band, from \STJ{}$\rightarrow$\STJR{}, is, nevertheless,
highly satisfactory, particularly when considered in the context
of earlier works on \MINLOPRIME{}~%
\cite{Hamilton:2012rf,Luisoni:2013kna,Hamilton:2016bfu}.

The top quark transverse momentum distribution, in the
right-hand plot of Fig.~\ref{fig:top}, shows a similarly
expected and pleasing pattern of results.
As for $y(t)$, there is a very small scale uncertainty
associated to the NLO accurate \ST{} predictions for
this observable, not exceeding $\pm 4\%$.
The central scale, LO accurate \STJ{} prediction --- which is simply
divergent without the \MINLO{} prescription of
section~\ref{subsec:STJ-MiNLO} --- lies within 10\% of the
central \ST{} one below $p_{\mathrm{T}}(t) = 1\,\mathrm{TeV}$.
Again, the \STJ{} result exhibits a relatively large
uncertainty band, consistent with that seen in the
$y(t)$ distribution. By contrast, the \STJR{} prediction
sits within $\pm 2\%$ of the NLO accurate \ST{} result
all through the range
$10 < p_{\mathrm{T}}(t) < 1250\,\mathrm{GeV}$. %
For $p_{\mathrm{T}}(t) < 10\, \mathrm{GeV}$ the \STJR{}
prediction deviates by up to 7\% from the central \ST{}
prediction. However, the cross section is falling very
steeply in this part of the spectrum, reducing by a
factor of $\sim 5$, in the interval
$5 < p_{\mathrm{T}}(t) < 10\,\mathrm{GeV}$.

\subsection{Inclusive jet cross sections}
\label{subsec:inclusive-jet-xsecns}

\begin{figure}[t]
\begin{center}  
  \includegraphics[width=0.459\textwidth]%
                  {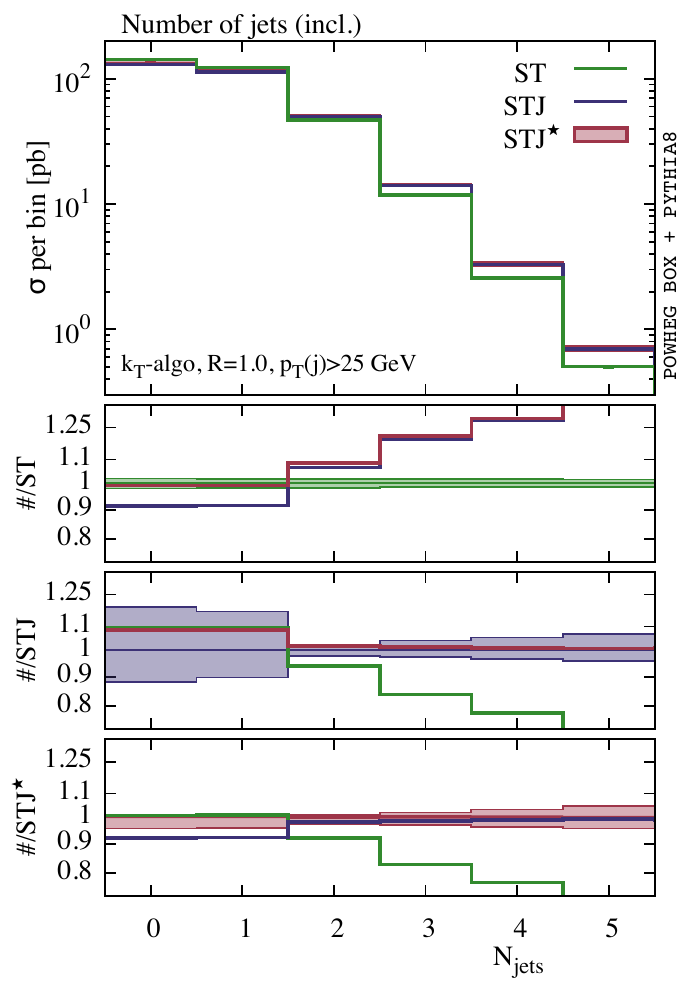}
  \hspace{10 mm}
  \includegraphics[width=0.459\textwidth]%
                  {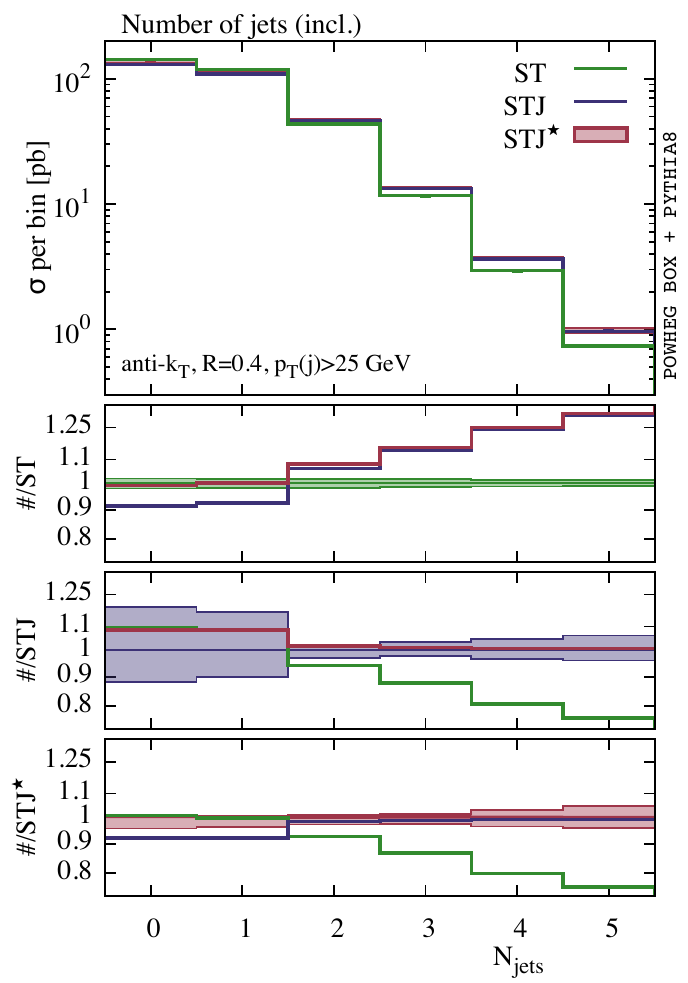}
\end{center}  
\caption{Inclusive jet cross sections in $t$-channel
  single-top production, with a jet transverse momentum
  threshold of 25 GeV. The left-hand plot shows predictions
  for jets defined according to the $k_{t}$ clustering
  algorithm with radius parameter $R = 1$, while the
  right-hand plot gives the analogous predictions for
  the case of the anti-$k_{t}$ algorithm with $R = 0.4$.
  As in Fig.~\ref{fig:top} we show in green, blue and red,
  predictions from the \ST{}, \STJ{} and \STJR{} simulations
  respectively.
  \label{fig:sigincl}
  }
\end{figure}

Fig.~\ref{fig:sigincl} shows the inclusive jet cross sections
in $t$-channel single-top production, for jets formed by the
radius $R=1$ $k_{t}$ algorithm, on the left, and the $R=0.4$
anti-$k_{t}$ algorithm, on the right. In both cases a transverse
momentum cut of $25$ GeV is applied in defining the jets.
The $R=1$ inclusive $k_{t}$ jet cross sections are primarily
of technical interest, being the inclusive version of the
jet definition used in tuning the \STJ{} \MINLO{} Sudakov form
factor. The $R=0.4$ anti-$k_{t}$ jet cross sections are more
experimentally relevant, since this is the typical jet definition
employed in LHC analyses.

For both jet definitions the results shown in
Fig.~\ref{fig:sigincl} are very much just as we would like.
The ${\mathrm N_{\mathrm{jets}}}\ge 0$ and ${\mathrm N_{\mathrm{jets}}}\ge 1$
cross sections are inclusive $t$-channel single-top production
observables, receiving their leading contributions in perturbation
theory from the lowest order $bq \rightarrow tq^{\prime}$ process.
Accordingly, the \ST{} predictions (green) are NLO accurate in
describing these jet bins. Conventional fixed order $t$-channel
single-top plus jet predictions for the same cross sections would
be divergent. In contrast, through inclusion of the
\MINLO{} procedure (section~\ref{subsec:STJ-MiNLO}), the
predictions of the \STJ{} simulation (blue) lie just
$\sim 10\%$ below those of the NLO accurate \ST{} results.
The central prediction of the \STJR{} code, with the tuned
\MINLO{} Sudakov form factor (red), further improves on the
latter, and converges exactly onto the NLO \ST{} predictions
in the same two jet bins.

Looking to the higher multiplicity cross sections, for
$N_{\mathrm{jets}}\ge 2$ we see the \STJR{} generator now
exactly aligns with the \STJ{} predictions, as opposed
to those of \ST{}. The \ST{} cross sections fall below
those of \STJ{} and \STJR{} by an amount which increases
with $N_{\mathrm{jets}}$. Both the \STJR{} and \STJ{}
predictions in the $N_{\mathrm{jets}}\ge 2$ and
$N_{\mathrm{jets}}\ge 3$ bins, are NLO and LO accurate
respectively. On the other hand, in the \ST{} case, the
description of $N_{\mathrm{jets}}\ge 2$ is LO accurate, while
events in the $N_{\mathrm{jets}}\ge 3$ bin are due entirely
to parton showering. The undershooting of jet cross sections
by simulations based on lower multiplicity matrix elements,
compared to those built from higher multiplicity ones, is
a typical observation in comparisons of event generators
based on matrix element-parton shower matching/merging.

\subsection{Differential jet rates}
\label{subsec:differential-jet-rates}

\begin{figure}
\begin{center}  
  \includegraphics[width=0.459\textwidth]%
                  {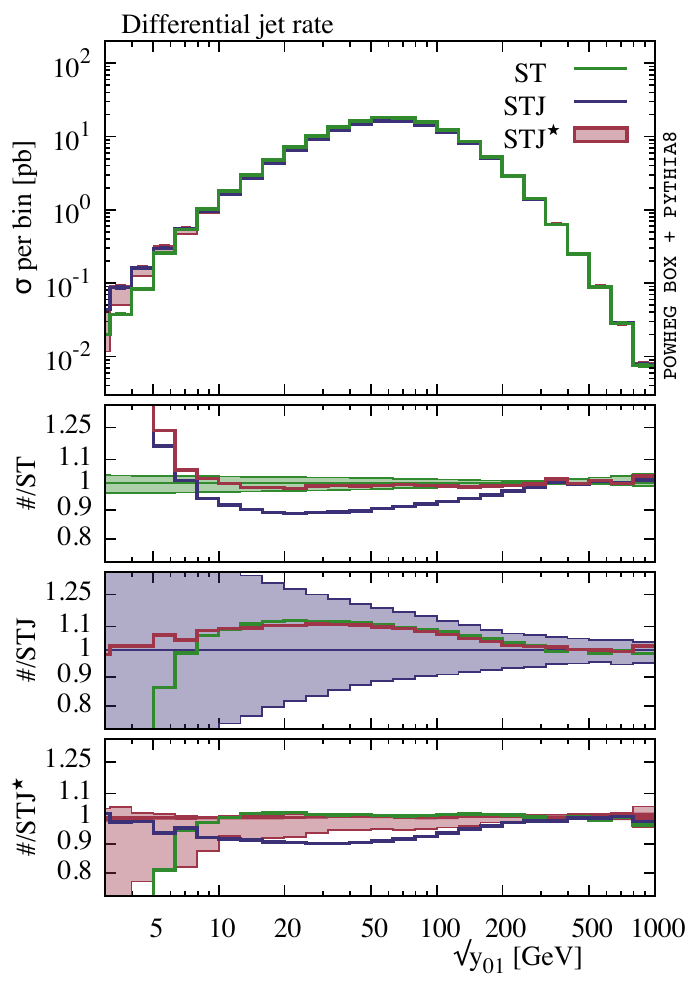}  
  \hspace{10 mm}
  \includegraphics[width=0.459\textwidth]%
                  {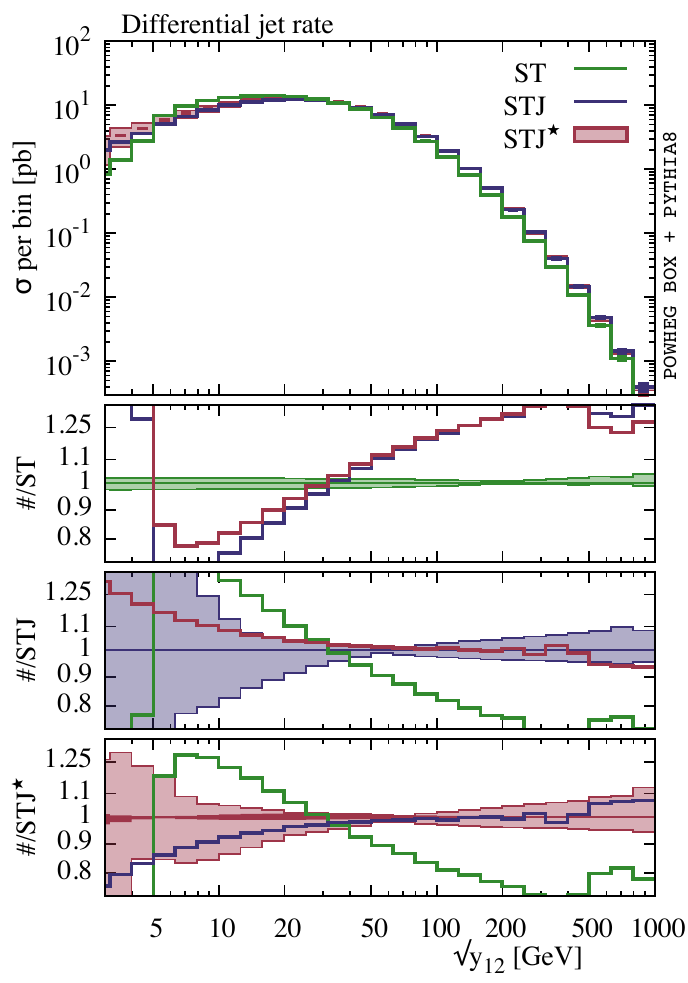}
\end{center}  
\caption{Differential jet rates in the exclusive $k_{t}$
  clustering algorithm~\cite{Catani:1993hr}, with
  jet radius parameter $R=1$. The left-hand
  plot presents predictions for the $0 \rightarrow 1$ jet
  rate, $\sqrt{y_{01}}$, corresponding to the value of the
  distance measure in that algorithm at which a 1-jet event
  would become resolved as a 0-jet one. The right-hand plot
  shows the $1 \rightarrow 2$ jet rate, $\sqrt{y_{12}}$,
  analogously defined. As in Figs.~\ref{fig:top}
  and \ref{fig:sigincl}, all predictions include the effects
  of parton showering provided by \PYTHIAEIGHT{}, and 
  follow the same colour conventions.  
}
\label{fig:diffjetrates}
\end{figure}

The $n \rightarrow m$ differential jet rates, $y_{nm}$,
measure the value of the distance measure in the exclusive
$k_{t}$ clustering algorithm at which an $n$-jet event
becomes resolved as an $m$-jet one. They are key variables
of interest in validating our \STJ{} and \STJR{} generators.

The $\sqrt{y_{01}}$ jet rate, on the left-hand side of
Fig.~\ref{fig:diffjetrates}, is essentially equivalent
to the transverse momentum spectrum of the hardest jet
obtained in the inclusive $k_{t}$ clustering algorithm,
with jet radius $R=1$.
Hence $\sqrt{y_{01}}$ is therefore described with NLO
accuracy by the \ST{} simulation and LO accuracy by \STJ{}.
Correspondingly, except for the region
$\sqrt{y_{01}}\lesssim 5$~GeV, the blue \STJ{} prediction
lies within $\sim 10\%$ of the green \ST{} result. In
the same region, all the way up to $\sqrt{y_{01}} = 1$~TeV
the central \STJR{} prediction lies within the tiny \ST{}
scale uncertainty band, which is never more than $\pm 4\%$
wide. Moreover, the \STJR{} scale uncertainty band is,
again, greatly shrunk with respect to that of the \STJ{}
simulation, being at the level of $+2\%$/$-6\%$ down to
$\sqrt{y_{01}}\lesssim 10$~GeV. This level of agreement
is satisfying considering that a linear plot of the leading
jet transverse momentum spectrum (not shown) reveals that
the cross section falls by five orders of magnitude in
the interval $10 \rightarrow 1000$~GeV.

As we approach 5 GeV in the $\sqrt{y_{01}}$ spectrum from
above, we observe a sharp irregular behaviour from the
NLO accurate \ST{} generator. In particular, the latter
distribution exhibits a sharp downward step with respect
to the \STJ{} and \STJR{} predictions. This same trend is
also clear very close to 5 GeV in the transverse momentum
spectra of the first and second jets (not shown). The
feature arises due to the fact that the \ST{} program
generates real radiation events from $bq \rightarrow tq^{\prime}$
underlying Born configurations via the \POWHEG{} Sudakov
form factor. The latter Sudakov form factor exponent
contains $b$-quark PDFs in its numerator and denominator,
evaluated at the transverse momentum scale of the would-be
emitted radiation, $p_{\scriptscriptstyle{\mathrm{T,rad}}}$.
The $b$-quark PDFs evaluate to zero as soon as
$p_{\scriptscriptstyle{\mathrm{T,rad}}}<m_{b,0}$, where $m_{b,0}$
is the value of the factorization scale at which the
$b$-quark density is turned on or off in the relevant
PDF set. Finally, we stress that the significance of these irregularities
in the differential jet rates and jet transverse
momentum spectra should not be overstated, since they occur
only at low scales that are of limited phenomenological and
experimental relevance.

We now turn our attention to the $\sqrt{y_{12}}$ distribution
shown on the right-hand side of Fig.~\ref{fig:diffjetrates}.
This distribution is both very important and informative in
checking the effects due to the \MINLO{} tuning procedure,
since it is precisely this quantity, albeit defined at the
level of pre-shower Les Houches events, which the tuning acts
on directly. This distribution therefore measures very well
the cost, or any potential breakage, associated with promoting
\STJ{}$\rightarrow$\STJR{}.

Away from the Sudakov peak region, $\sqrt{y_{12}}\gtrsim 20$~GeV,
where it is meaningful to talk of accuracy defined in terms
of fixed order perturbation theory, the \ST{} simulation is
only LO accurate, while \STJ{} is NLO accurate. As expected,
we see that the \STJR{} simulation, which fully aligns with
the \ST{} predictions above 10~GeV in the $\sqrt{y_{01}}$
spectrum, here, instead, agrees with \STJ{} to the
right of the Sudakov peak in $\sqrt{y_{12}}$. Below
$\sqrt{y_{12}}=30$~GeV the central \STJ{} and \STJR{} results
begin to slowly deviate from one another, due to the effects
of the tuning in the latter's \MINLO{} Sudakov form factor;
the differences reach 3-4\% at $\sqrt{y_{12}} = 20$~GeV,
rising to 8\% deep in the Sudakov region at
$\sqrt{y_{12}} = 10$~GeV, with neither prediction ever
departing outside the other one's scale uncertainty band. We
can see from this plot that the relatively low \STJ{} cross
sections observed for inclusive quantities and low multiplicity
jet cross sections, in sections~\ref{subsec:top-rapidity-and-pT}-%
\ref{subsec:inclusive-jet-xsecns}, are compensated for in the
\STJR{} simulation by the uplift in its $\sqrt{y_{12}}$ spectrum
with respect to that of \STJ{}, both on and below the Sudakov
peak  in the $\sqrt{y_{12}}$ distribution. It is further clear
from this spectrum that the NLO accuracy of \STJ{} for $t$-channel
single-top plus jet observables has been fully inherited by
the tuned \STJR{} simulation.

Finally, we remark that the same smallness of the \ST{}
uncertainty band seen in the $\sqrt{y_{01}}$ distribution
persists in the $\sqrt{y_{12}}$ spectrum and is, again,
an underestimate of the true uncertainty.
It is an artefact of the
\POWHEG{} \NLOPS{} methodology, whereby the scale compensation
associated with NLO accurate $bq \rightarrow tq^{\prime}$
underlying Born kinematics is spread out all through the
single-top plus jet phase space. The \ST{} uncertainty
is also underestimated in the region below the peak at
$10$~GeV, which is dominated by large Sudakov logarithms
at all orders and by non-perturbative effects.
The \STJ{} and
\STJR{} simulations are, conversely, NLO rather than LO for
this distribution, and while they carry larger uncertainty
bands than \ST{}, their estimates should be
considered to be much more realistic for the $\sqrt{y_{12}}$
spectrum.

\subsection{Top-jet angular correlations}
\label{subsec:top-jet-angular-correlations}

Angular correlations between the top quark and the leading jet are
somewhat complementary to the $\sqrt{y_{12}}$ differential jet rate
just discussed, since they also probe across the transition between
topologies involving one and two resolved jets, albeit now in terms of
angles.

\begin{figure}[t]
\begin{center}  
  \includegraphics[width=0.459\textwidth]{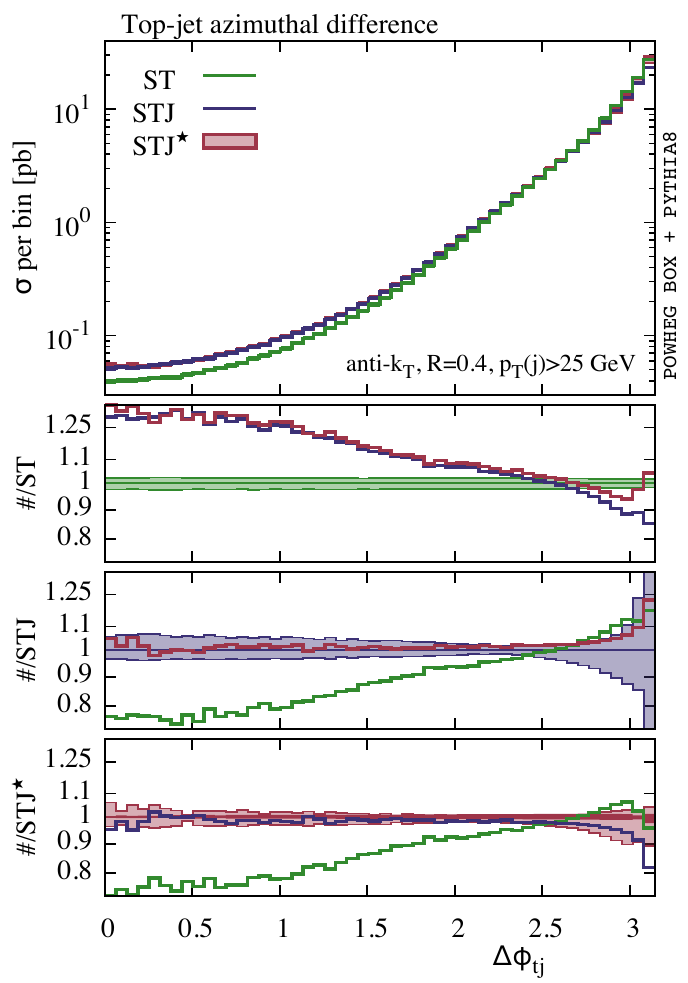}
  \hspace{10 mm}
  \includegraphics[width=0.459\textwidth]{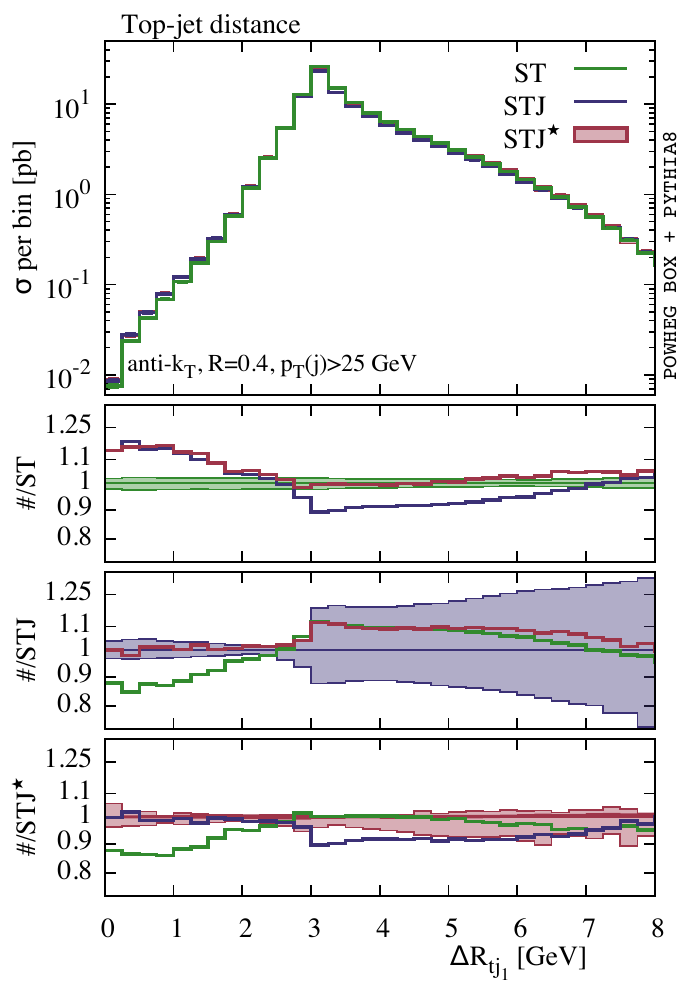}  
\end{center}  
\caption{
  The azimuthal separation of the top quark and the leading
  jet (left), and the $\eta-\phi$ plane distance,
  $\Delta R_{tj_{1}} = (\Delta \phi_{tj}^2 + \Delta \eta_{tj}^2)^{\frac{1}{2}}$,
  between the same two objects (right).
  Predictions from the \POWHEG{} \ST{} program~\cite{Alioli:2009je}
  are shown in green. Results from the new \MINLO{} \STJ{}
  simulation are displayed in blue, while those of its improved
  counterpart, \STJR{}, appear in red. All predictions include
  parton shower effects simulated by
  \PYTHIAEIGHT{}~\cite{Sjostrand:2014zea}.
}
\label{fig:DRtj}
\end{figure}

On the left of Fig.~\ref{fig:DRtj} we show the azimuthal
separation between the top quark and the leading jet in
$t$-channel single-top production. At lowest order in
perturbation theory this distribution would consist of a
lone spike at $\Delta \phi_{tj} = \pi$, since at that order
the top quark and the light parton must exactly balance each
other's transverse momentum. Additional soft-collinear radiation on
top of the latter smears the spike out into the peak seen
around $\Delta \phi_{tj} = \pi$, in the left-hand plot of
Fig.~\ref{fig:DRtj}. Furthermore, the integral of the
$\Delta \phi_{tj}$ distribution must, by definition, yield
the inclusive 1-jet cross section of Fig.~\ref{fig:sigincl}.
Thus, the normalizaton of this distribution, which is largely
set by the peak region, is described with NLO accuracy by
\ST{} and LO accuracy by \STJ{}. Taking the above points
together, it then makes sense that we see the \STJR{} program
tend to the \ST{} prediction in the peak region. Indeed, the
$\sim 10-15\%$ deficit between the \STJ{} prediction and
that of \ST{}, in the region $\Delta \phi_{tj} = \pi$, correlates
with the $\sim 10\%$ deficit seen in the inclusive 1-jet
cross section on the right of Fig.~\ref{fig:sigincl}. Equally,
the agreement of \ST{} and \STJR{} in the peak region is
reflective of the corresponding agreement in the inclusive
1-jet cross section of Fig.~\ref{fig:sigincl}.

Moving off the peak region in $\Delta \phi_{tj}$, the distribution
becomes increasingly populated by topologies involving the top
quark recoiling against two jets, or more. In fact the region
$\Delta \phi_{tj} \lesssim 2\pi/3$ is not accessible if the top quark
only recoils against two final-state objects. Correspondingly, off
the peak region, the \STJ{} simulation can be expected to give the
most accurate predictions (NLO).
Pleasingly, and expectedly, we see the \STJR{} simulation is
indistinguishable from the \STJ{} prediction already for
$\Delta \phi_{tj} \lesssim 2.6$.

The distribution on the right-hand side of Fig.~\ref{fig:DRtj} plots
the angular distance between the top quark and the leading jet in the
$\eta-\phi$ plane: $\Delta R_{tj} = (\Delta \phi_{tj}^2 + \Delta
\eta_{tj}^2)^{1/2}$.  The events populating the peak region in this
plot are predominantly those in the peak of the $\Delta \phi_{tj}$
distribution, albeit with the top quark and its balancing light jet both at
relatively central rapidities.
Also, the region to the right of the peak is
dominantly comprised of events with a top quark back-to-back in
azimuth with the leading light jet. In other words, the region
close to and above $\Delta R_{tj} = \pi $ is filled by events with
one resolved jet, and it is therefore described with NLO accuracy
by the \ST{} simulation, and LO accuracy by \STJ{}. Once again,
we see that the tuned \MINLO{} Sudakov form factor in the \STJR{}
simulation works as intended, with its prediction (red) falling
within the $\lesssim 3\%$ uncertainty band of \ST{} (green), and
exceeding the \ST{} central value by not more than $5\%$, across
the region $\Delta R_{tj} \gtrsim \pi$.

The only way to populate the $\Delta R_{tj}$ region to the left of the
peak is to have $\Delta \phi_{tj} < \pi$, moreover, the only way to
populate $\Delta R_{tj} \lesssim 2\pi/3$ is with events in which the
top quark recoils against more than two final-state objects. The
latter description is, of course, familiar from the discussion on the
$\Delta \phi_{tj}$ distribution just overhead, owing to the fact that,
by definition, $\Delta R_{tj} \ge \Delta \phi_{tj}$.
As a consequence, the
NLO \ST{} simulation is only LO accurate in the region $2\pi/3 \lesssim
\Delta R_{tj} \lesssim \pi$, while it relies on the parton shower to
fill $\Delta R_{tj} \lesssim 2\pi/3$. Conversely, the \STJ{} simulation
will be NLO accurate in the region $2\pi/3 \lesssim \Delta R_{tj}
\lesssim \pi$, and LO below it. Once again, \STJR{} is seen to behave
in the best possible way, moving away from the \ST{} prediction on the
peak at $\Delta R_{tj} \sim \pi$ and aligning exactly with the \STJ{}
result in the region below $\Delta R_{tj} \sim 2.6$.

Before moving on, we point out that the same $\Delta \phi_{tj}$ and
$\Delta R_{tj}$ distributions seen here in Fig.~\ref{fig:DRtj} are
reproduced for the 8 TeV LHC in appendix~\ref{app:Tune-extrapolation}
(Fig.~\ref{fig:DRtj_8tev}). These 8 TeV results display, quantitatively,
exactly the same trends as those shown here, in particular they show
the same excellent agreement between \STJR{} and \ST{}/\STJ{}
simulations in the same regions of the plots elaborated on above. We
emphasise that the 8 TeV \STJR{} predictions were
obtained using the same neural network fit of the \MINLO{} Sudakov
form factor as employed for the plots in this section,
suggesting that the fit comes with a reasonable degree of
portability/universality.

\subsection{Top transverse momentum in single jet events}
\label{subsec:top-transverse-momentum-in-single-jet-events}

\begin{figure}
\begin{center}  
  \includegraphics[width=0.459\textwidth]{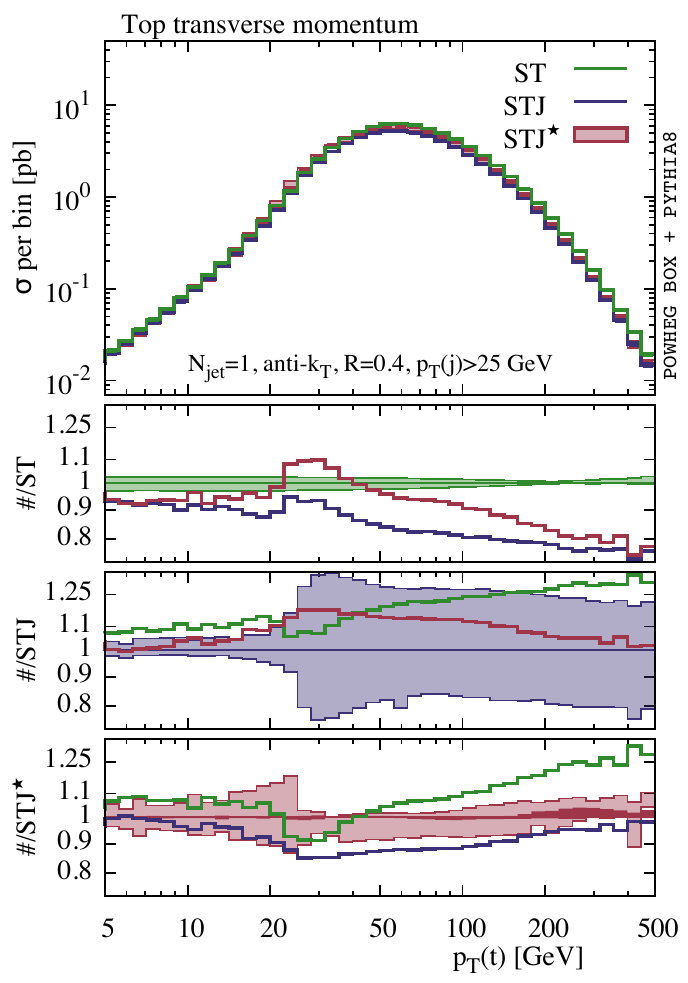}
  \hspace{10 mm}
  \includegraphics[width=0.459\textwidth]{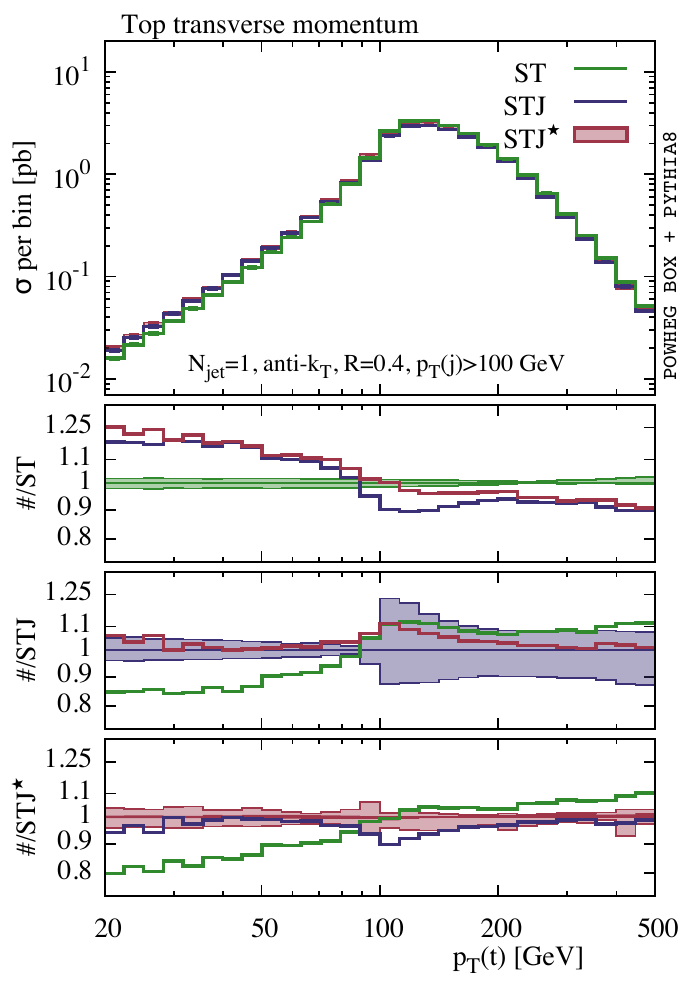}  
\end{center}
\caption{
  Transverse momentum of the top quark in events containing
  exactly one light $R=0.4$ anti-$k_{t}$ jet in addition to
  the top quark. On the left the distribution is defined with
  a $25$~GeV jet transverse momentum threshold, while on the
  right-hand side a $100$~GeV threshold is used. No other
  cuts are applied.  All predictions follow the same colour
  conventions as Figs.~\ref{fig:top}-\ref{fig:DRtj}.
 }
\label{fig:topptcuts}
\end{figure}

In Fig.~\ref{fig:topptcuts} we show how the top quark's
transverse momentum spectrum, Fig.~\ref{fig:top}, is
modified by requiring that it be accompanied by exactly one
light jet. The distribution is plotted for two different
jet transverse momentum thresholds: 25~GeV (left) and 100~GeV
(right). The spectrum is of interest since the requirement
to have only one light jet, in addition to the top quark, has
been employed in event selections, as a means to reduce
background, in LHC $t$-channel single-top
analysis~\cite{Aaboud:2017pdi}.

In both left- and right-hand plots of Fig.~\ref{fig:topptcuts},
in order to populate the region where the top quark's
$p_{\scriptscriptstyle{\mathrm{T}}}(t)$ is low with respect to
the transverse momentum threshold at which jets are defined,
and yet still have single-jet events, there must be a second
collimated spray of radiation, to balance the
transverse momentum. Hence, the events in this region are
expected to be two-jet like, with the resolved jet and its
would-be-jet counterpart nearly back-to-back in azimuth
as $p_{\scriptscriptstyle{\mathrm{T}}}(t)\rightarrow 0$~GeV.
It follows that this region of
$p_{\scriptscriptstyle{\mathrm{T}}}(t)$ is best described by
the \STJ{} program, with NLO accuracy, while the corresponding
description in the \ST{} program is LO. Being two-jet-like,
the \MINLO{} Sudakov form factor, tuned or not, will not act
on these events, and so it is natural then to see \STJR{}
(red) lie on top of \STJ{} (blue), as we approach 
$p_{\scriptscriptstyle{\mathrm{T}}}(t)=0$~GeV. The relative
closeness of the \ST{} prediction to the latter is consistent
with it having LO accuracy in the same region.\footnote{
  The smallness of the \ST{} uncertainty band here is again
  an underestimate of the true theory uncertainty. 
  It is a general feature in \POWHEG{} \NLOPS{}
  simulations, wherein the scale compensation associated
  with the NLO underlying Born kinematics, by default, is
  spread through all of the real radiation phase space.
}

We now turn, temporarily, to consider the left-hand plot,
wherein a 25~GeV jet transverse momentum threshold is in
effect. If one looks to the high
${\mathrm{p}}_{\scriptscriptstyle{\mathrm{T}}}(\mathrm{t})$
end, one is certainly considering a region dominated by
large Sudakov logarithms, of sizeable ratios of scales
$\sim Q_{bt}^{2}/p_{\scriptscriptstyle{\mathrm{T,cut}}}^{2}$
(eq.~\eqref{eq:sect22-Qbt-eq-two-pb-dot-pt}),
where $p_{\scriptscriptstyle{\mathrm{T,cut}}}$ denotes the jet
transverse momentum threshold, at which the presence of a
second radiated jet is vetoed. Thus, any variance (or lack
of it), in the predictions for this tail of the spectrum,
owes to differences in the Sudakov region around the
$bq \rightarrow tq^{\prime}$ underlying Born in each
simulation.

In the peak of the distribution,
${\mathrm{p}}_{\scriptscriptstyle{\mathrm{T}}}(\mathrm{t}) \sim 50$~GeV,
close to, but clearly above $p_{\scriptscriptstyle{\mathrm{T,cut}}}$,
the make-up of the predictions, and our expectations
for them, is less clear, owing to the complicated nature
of the observable and the various dynamics which enter. 
We tentatively suggest that the underlying Born configurations,
$bq \rightarrow tq^{\prime}$, associated with this region of the
spectrum, are such that the 25~GeV transverse momentum veto on
the presence of two or more jets, does not greatly restrict the
phase space for radiation from those states. Assuming this to
be the case, being inclusive w.r.t.~radiation emitted from the
underlying Born configurations, it is then not surprising to
see the \STJ{} prediction $\sim 10-20\%$ below the \ST{} one
in this region of the distribution. For the same reasons, the
relative agreement of \STJR{} and \ST{} in the same vicinity,
is also anticipated, and desirable. 

We suggest the same tentative explanations for the behaviour
shown in the peak region of the right-hand
${\mathrm{p}}_{\scriptscriptstyle{\mathrm{T}}}(\mathrm{t})$ spectrum,
as for that on the left. For the right-hand plot, however, the
larger jet transverse momentum threshold of 100~GeV, means that
the degree of integration over the phase space for additional
radiation, at any given point in this region of the
${\mathrm{p}}_{\scriptscriptstyle{\mathrm{T}}}(\mathrm{t})$ spectrum,
is more inclusive than in the case of the 25~GeV cut used
for the left-hand plot. Hence, the \STJR{} prediction appears
to follow that of \ST{} over a slightly longer interval in the
central region of the right-hand plot.

\section{Conclusions}\label{sec:conclusions}

In this work we have developed a new NLO accurate simulation of
$t$-channel single-top plus jet production, with matching to
parton showers via the \POWHEG{} method. The calculation has
been carried out in the structure function approximation, 
wherein each of the fermion lines connected to the exchanged
$W$-boson, and their associated radiative corrections, are
treated as if they originated from two independent copies of
the QCD sector
(sections~\ref{subsec:definition}-\ref{subsec:NLOPS-construction}).

We have enhanced the NLO calculation underlying the simulation
by applying a process-specific formulation of the \MINLO{}
method, as set out in section~\ref{subsec:STJ-MiNLO}. The
resulting \STJ{} simulation yields NLO accuracy for $t$-channel
single-top plus jet observables, and LO accuracy for inclusive
$t$-channel single-top quantities
(appendix~\ref{subsec:Resummed-y12-jet-rate}-%
\ref{subsec:STJ-MiNLO-predictions-for-ST-observables}).

As well as producing a novel simulation for an important hadron
collider process, our efforts have also concentrated significantly
on the more general aim of improving and extending the \MINLO{}
method. To this end we have substantially evolved the proposal
of ref.~\cite{Frederix:2015fyz}. The latter article suggests
that \MINLO{} simulations can be made NLO accurate in describing
both the original process on which the simulation was based, as
well as that with one less jet, by fitting, approximately, unknown
higher order terms in the \MINLO{} Sudakov form factor to that
effect. In this work we have applied the same idea. We postulate
that the leading differences between the LO predictions of our
\STJ{} simulation and conventional NLO, for inclusive $t$-channel
single-top production, owe to $\mathrm{NNLL}_{\sigma}$ terms in
the \MINLO{} Sudakov form factor that we do not control
(section~\ref{subsec:Improved-STJ-MiNLO}).
We significantly improve on ref.~\cite{Frederix:2015fyz} by
fitting such an $\mathrm{NNLL}_{\sigma}$ correction to the Sudakov
form factor directly in its exponent, rather than in its expansion.
At the same time, we employ a more refined methodology in
performing the fit, making use of advanced machine learning
techniques for this purpose (section~\ref{sec:NN}). The neural
network machinery used in this part of our construction makes
no assumptions regarding any dependence that the correction
to the \MINLO{} Sudakov form factor may have on the underlying
Born kinematics, $bq \rightarrow tq^{\prime}$, and it is given in terms 
of just 42 parameters, including weights and biases. 

Our machine learning framework was applied to determine an
approximate $\mathrm{NNLL}_{\sigma}$ correction to the \MINLO{}
Sudakov form factor using $\mathcal{O}(20\mathrm{M})$ \ST{} and
\STJ{} simulation events, produced for a 13 TeV LHC setup. The
latter fitted term in our simulation is implemented as a small
overall multiplicative correction to the weights of the events
in the \STJ{} Les Houches event files, which can be applied very
quickly. We refer to this `tuned' \STJ{} simulation output,
including the latter correction, as \STJR{}. The fit is performed
seven times for correlated, factor-two, variations of the
renormalization and factorization scales in the \ST{} and \STJ{}
generators. In this way the \STJR{} predictions for inclusive
$t$-channel single-top production yield very similar uncertainty
estimates to those of the \ST{} program, besides its central
prediction.

The \STJ{} and \STJR{} simulations were validated by comparing
them to one another, and to the pre-existing \ST{} \POWHEG{} program,
for $\mathcal{O}(200)$ observables, at a 13 TeV LHC. This validation
confirms well that the \STJ{} predictions are LO accurate for inclusive
$t$-channel single-top production, and NLO for the same process
with an additional jet. The results also confirm that the improved
\STJR{} simulation output is simultaneously NLO accurate for inclusive
$t$-channel single-top and single-top plus jet processes. A
representative selection of the distributions studied in our
validation have been presented and discussed in
section~\ref{sec:results}.

We have also carried out the same extensive analysis of observables
assuming an 8 TeV LHC setup. For the latter we produced new \ST{},
\STJ{} and \STJR{} event samples accordingly. While it only takes 8-10
hours on a single CPU to generate a new Sudakov form factor fit for
each scale choice, here in the \STJR{} simulation we continued to use
the same fit obtained from 13 TeV LHC events, in order to test its
universality.  We find, again, that the \STJR{} predictions reproduce
well the NLO accuracy of the \ST{} program for inclusive $t$-channel
single-top production observables. The latter observations are highly
suggestive of a robustness and universality in the tuned \MINLO{}
Sudakov form factor. This is as one would expect, if the leading
differences between the \ST{} and \STJ{} predictions for inclusive
quantities are, as postulated, mostly/fully accounted for by missing
higher order terms in the initial \MINLO{} Sudakov form factor.

We advise, however, that if a tuning is carried out for a given
collision energy and then used to reweight \STJ{} events
simulated at a much higher one, the enlarged $b q\rightarrow tq^{\prime}$
phase space of the latter requires the neural network model to
extrapolate outside the region covered by the data used to train
it, e.g. into regions with very high transverse momentum or very
high rapidity top quarks. One should therefore obviously not expect
the neural network tunes to work so well when simulating \STJ{}
production at hadronic centre-of-mass energies significantly
above those used in their training. For dedicated studies in
such circumstances a new tuning of the \MINLO{} Sudakov form
factor can be carried out.

Finally, we point out that, given an NNLO calculation for $t$-channel
single-top production, with the capability to compute distributions
differential in the $bq \rightarrow tq^{\prime}$ Born phase space, it
is straightforward, in principle, to develop an NNLOPS simulation of
this process using the methodology presented here.

The new \STJ{} generator, together with the corresponding fits for
promoting it to \STJR{}, will soon be publicly available in the
\POWHEGBOX{} V2 framework. While in this paper we have presented
results only for single-top production, the code and the fits will be
made available also for anti-top production.

\section*{Acknowledgements}

We are very grateful to Bernhard Mistlberger for numerous
thought-provoking discussions and encouragement throughout
the project. We thank Peter Skands for useful discussions
about the treatment of showering initial-final QCD dipoles
in \PYTHIAEIGHT{}. We also acknowledge helpful conversations
with Paolo Nason and Emanuele Re.
This work was supported in part by ERC Consolidator
Grant HICCUP (No.\ 614577). 
KH thanks CERN-TH for its kind hospitality in extended phases of
the project. KH also thanks the Science and Technology Facilities
Council (STFC) for support via grant award ST/P000274/1. 
RF and GZ thank MIAPP for hospitality while this work was being carried out.
RF is supported by the Alexander von Humboldt Foundation, in
the framework of the Sofja Kovaleskaja Award Project ``Event
Simulation for the Large Hadron Collider at High Precision'', endowed
by the German Federal Ministry of Education and Research.
The authors acknowledge the use of the UCL Legion High
Performance Computing Facility (Legion@UCL), and associated support
services, in the completion of this work.

\appendix

\section{\MINLO{} supplement\label{app:Supplementary-details-on-the-MiNLO-procedure}}

In this appendix we give additional explanations and insights regarding
key points of section~\ref{subsec:Improved-STJ-MiNLO}, where we described
how the accuracy of the new \STJ{} program can be extended to NLO for
\ST{} observables. To this end, as with previous works on improving
\MINLO{}~\cite{Hamilton:2012rf,Frederix:2015fyz} we compare the basic
\STJ{} cross section in section~\ref{subsec:STJ-MiNLO}, differential in
the underlying  \ST{} Born kinematics and the relevant radiation hardness
parameter, to an analogous resummation formula. In
appendix~\ref{subsec:Resummed-y12-jet-rate} we describe a matched, resummed,
cross section formula for the case at hand.
Appendix~\ref{subsec:STJ-MiNLO-predictions-for-ST-observables} compares
the basic \STJ{} simulation cross section to the latter, to clarify its
accuracy for \ST{} observables. Expanded explanation 
of our procedure for tuning the \MINLO{} Sudakov form factor, such that
the \STJ{} code provides NLO descriptions of both $t$-channel
single-top and $t$-channel single-top plus jet processes, is given
in appendix~\ref{subsec:Improving-STJ-MiNLO}.

\subsection{Resummation formula\label{subsec:Resummed-y12-jet-rate}}

Neglecting, momentarily, the top-quark mass, applying the \CAESAR{}
resummation formalism \cite{Banfi:2004yd}, the resummed cross section for
the $1\rightarrow2$ exclusive $k_{t}$-jet rate, $y_{12}$, in single-top
production, at next-to-leading log accuracy\footnote{Resummation of terms
of the form $\frac{1}{y_{12}}\,\bar{\alpha}_{{\scriptscriptstyle \mathrm{S}}}^{n}\,\ln^{m}\frac{Q}{y_{12}}$, with $m=2n-1$ and $m=2n-2$. } ($\mathrm{NLL}_{\sigma}$),
with matching to NLO fixed order perturbation theory, can be written as

\begin{align}
\frac{d\sigma_{{\scriptscriptstyle \mathcal{RF}}}}{d\Phi dy_{12}}\, & =\,\frac{d\sigma_{{\scriptscriptstyle \mathcal{R}}}}{d\Phi dy_{12}}\,+\,\frac{d\sigma_{{\scriptscriptstyle \mathcal{F}}}}{d\Phi dy_{12}}\,,\label{eq:sect222-dsigRF-defn}
\end{align}
where $d\sigma_{{\scriptscriptstyle \mathcal{F}}}$ is a fixed order
contribution, finite as $y_{12}\rightarrow0$, and $d\sigma_{{\scriptscriptstyle \mathcal{R}}}$
embodies the all-orders resummation:

\begin{eqnarray}
\frac{d\sigma_{{\scriptscriptstyle \mathcal{R}}}}{d\Phi dy_{12}} & = & \frac{d\sigma_{{\scriptscriptstyle \mathrm{LO}}}^{{\scriptscriptstyle \mathrm{ST}}}}{d\Phi}\,\left[1+\bar{\alpha}_{{\scriptscriptstyle \mathrm{S}}}\bar{\chi}_{1}\left(\Phi\right)\right]\,\frac{d}{dy_{12}}\,\left[\,\Delta(y_{{\scriptscriptstyle 12}})\,\prod_{\ell=1}^{n_{i}}\frac{q^{\left(\ell\right)}(x_{\ell},y_{12})}{q^{\left(\ell\right)}(x_{\ell},\mu_{{\scriptscriptstyle F}}^{2})}\,\right]\,.\label{eq:sect222-dsigR}
\end{eqnarray}
The first factor in eq.~\eqref{eq:sect222-dsigR}, $d\sigma_{{\scriptscriptstyle \mathrm{LO}}}^{{\scriptscriptstyle \mathrm{ST}}}/d\Phi$,
denotes the leading order cross section for single-top production,
fully differential in its associated kinematics, $\Phi$. The $\bar{\chi}_{1}\left(\Phi\right)$
term encodes hard virtual next-to-leading order corrections to $d\sigma_{{\scriptscriptstyle \mathrm{LO}}}^{{\scriptscriptstyle \mathrm{ST}}}/d\Phi$
such that 
\begin{align}
\frac{d\sigma_{{\scriptscriptstyle \mathrm{NLO}}}^{{\scriptscriptstyle \mathrm{ST}}}}{d\Phi}\, & =\,\frac{d\sigma_{{\scriptscriptstyle \mathrm{LO}}}^{{\scriptscriptstyle \mathrm{ST}}}}{d\Phi}\,\left[1+\bar{\alpha}_{{\scriptscriptstyle \mathrm{S}}}\bar{\chi}_{1}\left(\Phi\right)\right]+\int dy_{12}\,\frac{d\sigma_{{\scriptscriptstyle \mathcal{F}}}}{d\Phi dy_{12}}\,,\label{eq:sect222-chi-bar-defn}\\
\frac{d\sigma_{{\scriptscriptstyle \mathcal{F}}}}{d\Phi dy_{12}}\, & =\,\frac{d\sigma_{{\scriptscriptstyle \mathrm{LO}}}^{{\scriptscriptstyle \mathrm{STJ}}}}{d\Phi dy_{12}}-\left.\frac{d\sigma_{{\scriptscriptstyle \mathcal{R},1}}}{d\Phi dy_{12}}\right|_{\bar{\chi}_{1}\rightarrow0}\,,\label{eq:sect222-dsigF-defn}
\end{align}
where $d\sigma_{{\scriptscriptstyle \mathcal{R},1}}$ denotes the $\alpha_{{\scriptscriptstyle \mathrm{S}}}$ expansion of
$d\sigma_{{\scriptscriptstyle \mathcal{R}}}$.
The $q^{\left(\ell\right)}(x_{\ell},\mu^{2})$ factors are parton
distribution functions (PDFs), for a given incoming leg, $\ell$,
evaluated at momentum fraction $x_{\ell}$, and scale $\mu$. The
product of PDF ratios runs over $n_{i}=2$ incoming legs. Except for
the argument of $\alpha_{{\scriptscriptstyle \mathrm{S}}}$ in the
integrands of $\Delta(y_{{\scriptscriptstyle 12}})$, renormalization
and factorization scales are set to a hard scale characteristic of
the leading order single-top production process throughout eqs.~(\ref{eq:sect222-dsigRF-defn}-\ref{eq:sect222-dsigF-defn}).

In neglecting the top-quark mass to use the \CAESAR{} framework,
the various elements of eqs.~(\ref{eq:sect222-dsigR}-\ref{eq:sect222-dsigF-defn})
should be initially understood as defined in the $m_{t}\rightarrow0$
limit. Extrapolating eq.~\eqref{eq:sect222-dsigR} to include the finite
top-mass is then straightforward, involving no change to the form
of eqs.~(\ref{eq:sect222-dsigR}-\ref{eq:sect222-dsigF-defn}) but
rather just obvious extensions of the elements making them up.

Being in the final-state, the top quark, whether we neglect its mass
or not, does not affect the PDF dependence of eq.~\eqref{eq:sect222-dsigR}.
The Sudakov form factor exponent, on the other hand, must be supplemented
by a set terms which vanish in the $m_{t}\rightarrow0$ limit (eq.~\eqref{eq:sect22-Delta-bt-ii}),
i.e.~$\Delta(y_{{\scriptscriptstyle 12}})$ in eq.~\eqref{eq:sect222-dsigR}
should be hence understood as the full expression in eq.~\eqref{eq:sect22-Delta-eq-Delta-qqprime-Delta-bt}
rather than its $m_{t}\rightarrow0$ limit. We remind again that the latter
finite quark mass extension of the Sudakov form factor is identical
to that in eq.~10 of the $k_{t}$-jet rate resummation of ref.~\cite{Krauss:2003cr}.
The only other modification to the resummed expression in eq.~\eqref{eq:sect222-dsigR},
due to the finite top mass, is trivial, merely consisting of henceforth
understanding that $d\sigma_{{\scriptscriptstyle \mathrm{LO}}}^{{\scriptscriptstyle \mathrm{ST}}}$
and $d\sigma_{{\scriptscriptstyle \mathrm{NLO}}}^{{\scriptscriptstyle \mathrm{ST}}}$
refer to the LO and NLO single-top cross sections with the full top-mass
dependence.
The fixed order matching terms $\bar{\chi}_{1}$ and
$d\sigma_{{\scriptscriptstyle \mathcal{F}}}$ remain determined by
eq.~\eqref{eq:sect222-chi-bar-defn}, subject to the latter modifications.

\subsection{\STJ{} predictions for \ST{} observables\label{subsec:STJ-MiNLO-predictions-for-ST-observables}}

Here we compare the resummed and matched cross section of
eq.~\eqref{eq:sect222-dsigRF-defn} to that of \STJ{}, to better understand
its predictions for inclusive $t$-channel single-top observables, since we
know what these are in the case of eq.~\eqref{eq:sect222-dsigRF-defn}. To
this end we first recast $d\sigma_{{\scriptscriptstyle \mathcal{RF}}}$
(eq.~\eqref{eq:sect222-dsigRF-defn}) in the same form as the \STJ{} cross
section
$d\sigma_{{\scriptscriptstyle \mathrm{STJ}}}^{{\scriptscriptstyle \mathcal{M}}}$
(eq.~\eqref{eq:sect22-dsigma-M-in-terms-of-dsigma-NLO-etc}). With no
approximations we can rewrite $d\sigma_{{\scriptscriptstyle \mathcal{R}}}$
in eq.~\eqref{eq:sect222-dsigR} as

\begin{eqnarray}
  \frac{d\sigma_{{\scriptscriptstyle \mathcal{R}}}}{d\Phi dy_{12}} &%
  = &%
  \Delta(y_{{\scriptscriptstyle 12}})\,%
  \frac{d\sigma_{{\scriptscriptstyle \mathrm{LO}}}^{{\scriptscriptstyle \mathrm{ST}}}}%
       {d\Phi}\,%
  \left[%
    1+\bar{\alpha}_{{\scriptscriptstyle \mathrm{S}}}\bar{\chi}_{1}%
    \left(\Phi\right)%
  \right]\,%
  \frac{d}{dy_{12}}\ln%
  \left[\,%
    \Delta(y_{{\scriptscriptstyle 12}})\,%
    \prod_{\ell=1}^{n_{i}}q^{\left(\ell\right)}(x_{\ell},y_{12})\,%
  \right]\,%
  ,\label{eq:sect223-dsigR-rewritten-exactly}
\end{eqnarray}
wherein the renormalization and factorization scales are now set
to $\sqrt{y_{12}}$ throughout, save for those in the integrands of
$\Delta(y_{{\scriptscriptstyle 12}})$, which remain evaluated at $q$,
as set out in section~\ref{subsec:STJ-MiNLO}.
Neglecting $\mathcal{O}(\alpha_{{\scriptscriptstyle \mathrm{S}}}^{2})$
terms which are finite as $y_{12}\rightarrow0$, we introduce a factor
$\Delta(y_{{\scriptscriptstyle 12}})\,[1+\bar{\alpha}_{{\scriptscriptstyle \mathrm{S}}}\bar{\chi}_{1}]$
in front of $d\sigma_{{\scriptscriptstyle \mathcal{F}}}$ in
eq.~\eqref{eq:sect222-dsigRF-defn}, setting
$\mu_{{\scriptscriptstyle R}}=\mu_{{\scriptscriptstyle F}}=\sqrt{y_{12}}$
throughout that term. Taken together with
eq.~\eqref{eq:sect223-dsigR-rewritten-exactly}, this gives, via
eq.~\eqref{eq:sect222-dsigF-defn}, without further approximation,

\begin{align}
  \frac{d\sigma_{{\scriptscriptstyle \mathcal{RF}}}}{d\Phi dy_{12}}\, &%
  =\,%
  \Delta(y_{{\scriptscriptstyle 12}})\,%
  \left[%
    \frac{d\sigma_{{\scriptscriptstyle \mathrm{NLO}}}^{{\scriptscriptstyle \mathrm{APX}}}}{d\Phi dy_{12}}%
    -%
    \left.%
    \Delta(y_{{\scriptscriptstyle 12}})%
    \right|_{\bar{\alpha}_{{\scriptscriptstyle \mathrm{S}}}}%
    \frac{d\sigma_{{\scriptscriptstyle \mathrm{LO}}}^{{\scriptscriptstyle \mathrm{STJ}}}}{d\Phi dy_{12}}\,%
    \right]\,,%
  \label{eq:sect223-dsigRF-rewritten-in-MiNLO-style}\\
  \frac{d\sigma_{{\scriptscriptstyle \mathrm{NLO}}}^{{\scriptscriptstyle \mathrm{APX}}}}{d\Phi dy_{12}}\, &%
  =\,%
  \frac{d\sigma_{{\scriptscriptstyle \mathrm{LO}}}^{{\scriptscriptstyle \mathrm{STJ}}}}{d\Phi dy_{12}}\,%
  \left[%
    1%
    +%
    \left.%
    \Delta(y_{{\scriptscriptstyle 12}})%
    \right|_{\bar{\alpha}_{{\scriptscriptstyle \mathrm{S}}}}%
    +%
    \bar{\alpha}_{{\scriptscriptstyle \mathrm{S}}}\bar{\chi}_{1}\left(\Phi\right)\right]\,.%
  \label{eq:sect223-dsigAPXNLO}
\end{align}

Since the aforementioned neglected
$\mathcal{O}(\alpha_{{\scriptscriptstyle \mathrm{S}}}^{2})$ terms are finite
as $y_{12}\rightarrow0$, eq.~\eqref{eq:sect223-dsigRF-rewritten-in-MiNLO-style}
is completely unchanged with respect to eq.~\eqref{eq:sect222-dsigRF-defn}
in regards to the logarithmic terms $\propto1/y_{12}$, and so too
is its fixed order accuracy up to and including terms of
$\mathcal{O}(\alpha_{{\scriptscriptstyle \mathrm{S}}})$.
This is the case both in the cumulant cross section and the $y_{12}$ spectrum. This
means, in particular, that $d\sigma_{{\scriptscriptstyle \mathcal{RF}}}/d\Phi$
remains equal to $d\sigma_{{\scriptscriptstyle \mathrm{NLO}}}^{{\scriptscriptstyle \mathrm{ST}}}/d\Phi$
up to $\mathcal{O}(\alpha_{{\scriptscriptstyle \mathrm{S}}}^{2})$
unenhanced terms.

To ease comparison, we write again here the \STJ{} formula,
eq.~\eqref{eq:sect22-dsigma-M-in-terms-of-dsigma-NLO-etc}, differential
in $\Phi$ and $y_{12}$,

\begin{equation}
  \frac{d\sigma_{{\scriptscriptstyle \mathcal{M}}}}{d\Phi dy_{12}}%
  =%
  \Delta(y_{{\scriptscriptstyle 12}})\,%
  \left[\,%
    \frac{d\sigma_{{\scriptscriptstyle \mathrm{NLO}}}^{{\scriptscriptstyle \mathrm{STJ}}}}{d\Phi dy_{12}}%
    -%
    \left.%
    \Delta(y_{{\scriptscriptstyle 12}})%
    \right|_{\bar{\alpha}_{{\scriptscriptstyle \mathrm{S}}}}%
    \frac{d\sigma_{{\scriptscriptstyle \mathrm{LO}}}^{{\scriptscriptstyle \mathrm{STJ}}}}{d\Phi dy_{12}}\,%
    \right]\,.%
  \label{eq:sect223-dsigMinlo}
\end{equation}
The difference between eqs.~\eqref{eq:sect223-dsigRF-rewritten-in-MiNLO-style}
and \eqref{eq:sect223-dsigMinlo} is clearly limited to the first term
in each of the square brackets in $d\sigma_{{\scriptscriptstyle \mathrm{NLO}}}^{{\scriptscriptstyle \mathrm{APX}}}$
and $d\sigma_{{\scriptscriptstyle \mathrm{NLO}}}^{{\scriptscriptstyle \mathrm{STJ}}}$.
Now let's zoom in on this.

Suppressing, for brevity, the $d\Phi\,dy_{12}$'s, and dropping terms
beyond $\mathrm{NLL}_{\sigma}$ accuracy, we can write $d\sigma_{{\scriptscriptstyle \mathcal{RF}}}\,=\,\Delta(y_{{\scriptscriptstyle 12}})\,d\sigma_{{\scriptscriptstyle \mathrm{LO}}}^{{\scriptscriptstyle \mathrm{STJ}}}$,
whereupon it follows that exactly to $\mathcal{O}(\alpha_{{\scriptscriptstyle \mathrm{S}}})$,
and to $\mathrm{NLL}_{\sigma}$ accuracy at $\mathcal{O}(\alpha_{{\scriptscriptstyle \mathrm{S}}}^{2})$,
$d\sigma_{{\scriptscriptstyle \mathrm{NLO}}}^{{\scriptscriptstyle \mathrm{APX}}}$
is the same as $d\sigma_{{\scriptscriptstyle \mathrm{NLO}}}^{{\scriptscriptstyle \mathrm{STJ}}}$,
i.e. 
\begin{equation}
d\sigma_{{\scriptscriptstyle \mathrm{NLO}}}^{{\scriptscriptstyle \mathrm{STJ}}}=d\sigma_{{\scriptscriptstyle \mathrm{NLO}}}^{{\scriptscriptstyle \mathrm{APX}}}+d\sigma_{{\scriptscriptstyle \mathrm{NLO}}}^{{\scriptscriptstyle \mathrm{RES}}}\,,\qquad d\sigma_{{\scriptscriptstyle \mathrm{NLO}}}^{{\scriptscriptstyle \mathrm{RES}}}=\frac{d\sigma_{{\scriptscriptstyle \mathrm{LO}}}^{{\scriptscriptstyle \mathrm{STJ}}}}{d\Phi dy_{12}}\,\bar{\alpha}_{{\scriptscriptstyle \mathrm{S}}}\,\mathcal{C}_{21}+\mathcal{O}(\alpha_{{\scriptscriptstyle \mathrm{S}}}^{2}/y_{12})\,,\label{eq:sect223-dsigSTJNLO-eq-dsigAPXNLO-plus-dsigRESNLO}
\end{equation}
where $\mathcal{C}_{21}$ is a $\Phi$-dependent $\mathcal{O}(1)$
coefficient that we do not presume to know. Inserting eq.~\eqref{eq:sect223-dsigSTJNLO-eq-dsigAPXNLO-plus-dsigRESNLO}
into eq.~\eqref{eq:sect223-dsigMinlo} and integrating over $y_{12}$
yields 
\begin{equation}
\frac{d\sigma_{{\scriptscriptstyle \mathcal{M}}}}{d\Phi}\,-\,\frac{d\sigma_{{\scriptscriptstyle \mathcal{RF}}}}{d\Phi}\,=\,\frac{d\sigma_{{\scriptscriptstyle \mathrm{LO}}}^{{\scriptscriptstyle \mathrm{ST}}}}{d\Phi}\cdot\mathcal{O}(\alpha_{{\scriptscriptstyle \mathrm{S}}})\,,\label{eq:sect223-dsigRF-eq-dsigMiNLO-plus-order-alphaS}
\end{equation}
with the $\mathcal{O}(\alpha_{{\scriptscriptstyle \mathrm{S}}})$
ambiguity due to the leading (unknown) $\mathrm{NNLL}_{\sigma}$ term
in $\Delta(y_{{\scriptscriptstyle 12}})\,d\sigma_{{\scriptscriptstyle \mathrm{NLO}}}^{{\scriptscriptstyle \mathrm{RES}}}$.
Since $d\sigma_{{\scriptscriptstyle \mathcal{RF}}}/d\Phi$ is NLO
accurate, eq.~\eqref{eq:sect223-dsigRF-eq-dsigMiNLO-plus-order-alphaS}
means that the standard \STJ{} calculation in section~\ref{subsec:STJ-MiNLO},
has only LO accuracy for inclusive $t$-channel single-top observables. The
numerical comparisons in section~\ref{sec:results}, between \POWHEG{}
\ST{} and \STJ{} simulations, give strong numerical support to the
analysis here.

\subsection{\STJ{} $\rightarrow$ \STJR{}\label{subsec:Improving-STJ-MiNLO}}

While we do not in general control $\mathrm{NNLL}_{\sigma}$ terms, it's clear
that $d\sigma_{{\scriptscriptstyle \mathrm{NLO}}}^{{\scriptscriptstyle \mathrm{STJ}}}$
also includes the process-dependent
$\bar{\alpha}_{{\scriptscriptstyle \mathrm{S}}}\bar{\chi}_{1}\left(\Phi\right)$
term of
$d\sigma_{{\scriptscriptstyle \mathrm{NLO}}}^{{\scriptscriptstyle \mathrm{APX}}}$,
owing to the \ST{} hard virtual corrections implicit in the soft-collinear
limit of the NLO \STJ{} cross section. With the latter point in mind,
we postulate that any $\mathrm{NNLL}_{\sigma}$ extension of the resummed,
matched, cross section formula, $d\sigma_{{\scriptscriptstyle \mathcal{RF}}}$,
in section~\ref{subsec:Resummed-y12-jet-rate}, would have exactly the
same form as in eq.~\eqref{eq:sect222-dsigRF-defn} (or be may be
re-expressed as such), with the only difference being the inclusion
of $\mathrm{NNLL}_{\sigma}$ terms in the Sudakov form factor exponent.
The latter modification is exactly what previous works on improving \MINLO{}
in the context of other processes would advocate~\cite{Hamilton:2012rf,%
  Luisoni:2013kna,Frederix:2015fyz,Hamilton:2016bfu},
as well as the general \CAESAR{} resummation formalism for processes
involving only massless partons.

In the presence of such a modification the fixed order properties
of $d\sigma_{{\scriptscriptstyle \mathcal{RF}}}$ are unaltered. The
$\mathcal{O}(\alpha_{{\scriptscriptstyle \mathrm{S}}})$ radiation
spectrum of eq.~\eqref{eq:sect222-dsigRF-defn} with respect to the
\ST{} Born kinematics is unchanged by the introduction of
$\mathcal{O}(\alpha_{{\scriptscriptstyle \mathrm{S}}}^{2})$ terms
in the Sudakov exponent, and it is trivially still the case
that
$d\sigma_{{\scriptscriptstyle \mathcal{RF}}}/d\Phi%
=%
d\sigma_{{\scriptscriptstyle \mathrm{NLO}}}^{{\scriptscriptstyle \mathrm{ST}}}/d\Phi$.
The form of the \STJ{} cross section is also completely unchanged
with respect to eqs.~\eqref{eq:sect22-dsigma-M-in-terms-of-dsigma-NLO-etc} 
and \eqref{eq:sect223-dsigMinlo}.

Since the \STJ{} cross section is accurate to
$\mathcal{O}(\alpha_{{\scriptscriptstyle \mathrm{S}}}^{2})$ in the
$y_{12}$ spectrum, if the resummation formula with the modified
Sudakov form factor is $\mathrm{NNLL}_{\sigma}$ accurate, then it
must reproduce all $\mathrm{NNLL}_{\sigma}$ terms in the latter on
expansion in $\alpha_{{\scriptscriptstyle \mathrm{S}}}$. It follows
that the result of such a change in $\Delta(y_{12})$ in 
section~\ref{subsec:STJ-MiNLO-predictions-for-ST-observables}, is to
reduce the residual difference between
$d\sigma_{{\scriptscriptstyle \mathrm{NLO}}}^{{\scriptscriptstyle \mathrm{STJ}}}$
and its counterpart,
$d\sigma_{{\scriptscriptstyle \mathrm{NLO}}}^{{\scriptscriptstyle \mathrm{APX}}}$,
in the resummation formula,
eq.~\eqref{eq:sect223-dsigRF-rewritten-in-MiNLO-style}:
\begin{align}
  \frac{d\sigma_{{\scriptscriptstyle \mathrm{NLO}}}^{{\scriptscriptstyle \mathrm{APX}}}}%
       {d\Phi dy_{12}} & \,%
  \rightarrow\,%
  \frac{d\sigma_{{\scriptscriptstyle \mathrm{LO}}}^{{\scriptscriptstyle \mathrm{STJ}}}}%
       {d\Phi dy_{12}}\,%
  \left[%
    1+%
    \left.%
    \Delta(y_{{\scriptscriptstyle 12}})%
    \right|_{\bar{\alpha}_{{\scriptscriptstyle \mathrm{S}}}}%
    +%
    \bar{\alpha}_{{\scriptscriptstyle \mathrm{S}}}%
    \left[%
      \bar{\chi}_{1}\left(\Phi\right)%
      +%
      \mathcal{C}_{21}\left(\Phi\right)\right]\right]\,%
  ,\label{eq:sect224-dsigAPXNLO-redefinition}\\
  d\sigma_{{\scriptscriptstyle \mathrm{NLO}}}^{{\scriptscriptstyle \mathrm{RES}}} & \,%
  \rightarrow\,%
  \frac{d\sigma_{{\scriptscriptstyle \mathrm{LO}}}^{{\scriptscriptstyle \mathrm{ST}}}}%
       {d\Phi}\,%
  \frac{\mathcal{C}_{20}\left(\Phi\right)}%
       {y_{12}}%
  \bar{\alpha}_{{\scriptscriptstyle \mathrm{S}}}^{2}%
  +%
  \mathcal{O}(\alpha_{{\scriptscriptstyle \mathrm{S}}}^{2})\,.%
  \label{eq:sect224-dsigRESNLO-redefinition}
\end{align}
Combining eqs.~(\ref{eq:sect224-dsigAPXNLO-redefinition}-%
\ref{eq:sect224-dsigRESNLO-redefinition}) with
eqs.~\eqref{eq:sect223-dsigRF-rewritten-in-MiNLO-style},
\eqref{eq:sect223-dsigMinlo},
\eqref{eq:sect223-dsigSTJNLO-eq-dsigAPXNLO-plus-dsigRESNLO}
and integrating over $y_{12}$ then quickly yields
\begin{equation}
  \frac{d\sigma_{{\scriptscriptstyle \mathcal{M}}}}%
       {d\Phi}\,%
 -\,%
 \frac{d\sigma_{{\scriptscriptstyle \mathcal{RF}}}}%
      {d\Phi}\,%
 =\,%
 \frac{d\sigma_{{\scriptscriptstyle \mathrm{LO}}}^{{\scriptscriptstyle \mathrm{ST}}}}%
      {d\Phi}\cdot\mathcal{O}(\alpha_{{\scriptscriptstyle \mathrm{S}}}^{3/2})%
 \,.%
      \label{eq:sect224-dsigMiNLO-minus-dsigRF-eq-order-alphaS-1-point-5}
\end{equation}
From here it is then clear that if the coefficient of the suggested
$\mathrm{NNLL}_{\sigma}$ term in the Sudakov form factor was further
modified by a suitably defined, spurious, formally subleading term,
$\sim1+\mathcal{O}(\sqrt{\alpha_{{\scriptscriptstyle \mathrm{S}}}})$,
we can arrange that 
\begin{equation}
  \frac{d\sigma_{{\scriptscriptstyle \mathcal{M}}}}{d\Phi}\,%
  =\,%
  \frac{d\sigma_{{\scriptscriptstyle \mathcal{RF}}}}{d\Phi}\,%
  =\,%
  \frac{d\sigma_{{\scriptscriptstyle \mathrm{NLO}}}^{{\scriptscriptstyle \mathrm{ST}}}}%
       {d\Phi}%
  \,.%
  \label{eq:sect223-dsigMiNLO-eq-dsigRF-eq-dsigNLO}
\end{equation}
Assuming our postulate is valid, namely, that promoting
eq.~\eqref{eq:sect222-dsigRF-defn} from $\mathrm{NLL}_{\sigma}$ to
$\mathrm{NNLL}_{\sigma}$ accuracy amounts to including a missing term in
the Sudakov form factor of the form 
\begin{equation}
  \ln\delta\Delta(y_{12})\,%
  =\,%
  -\int_{y_{12}}^{Q_{bt}^{2}}%
  \frac{dq^{2}}{q^{2}}\,%
  \bar{\alpha}_{{\scriptscriptstyle \mathrm{S}}}^{2}\,%
  \mathcal{A}_{2}(\Phi)\,%
  \ln\frac{Q_{bt}^{2}}{q^{2}}\,,
  \label{eq:deltay12}
\end{equation}
then we may determine the unknown $\mathcal{A}_{2}(\Phi)$ therein,
up to a factor
$\sim1+\mathcal{O}(\sqrt{\alpha_{{\scriptscriptstyle \mathrm{S}}}})$,
by fitting it such that eq.~\eqref{eq:sect223-dsigMiNLO-eq-dsigRF-eq-dsigNLO}
is satisfied. The resulting improved \STJ{} cross section, \STJR{},
will then converge on the $\mathrm{NNLL}_{\sigma}$ resummation while
remaining NLO accurate for \STJ{} observables, and further acquiring
NLO accuracy for inclusive \ST{} ones.

While we assume a form for the resummation formula at
$\mathrm{NNLL}_{\sigma}$, we do not presume to know the details of
the related Sudakov ingredients at that order, so we allow 
the $\mathcal{A}_{2}$ coefficient to have a general dependence on $\Phi$
already for this reason alone. If the true $\mathrm{NNLL}_{\sigma}$
resummation turned out to be $\Phi$-independent, this should
be reflected by a relative flatness of the fitted $\mathcal{A}_{2}(\Phi)$.
However, even if this is the case at $\mathrm{NNLL}_{\sigma}$, the
relative $\mathcal{O}(\sqrt{\alpha_{{\scriptscriptstyle \mathrm{S}}}})$
ambiguity on the fitted $\mathcal{A}_{2}$, also absorbs the effects
of unknown $\mathrm{N^{3}LL}_{\sigma}$ Sudakov terms, which are established
as having a general dependence on the Born kinematics in so-called
process-dependent resummation formulae like that of \MINLO{}
(see e.g.~\cite{Catani:2000vq,Hamilton:2012rf,Hamilton:2016bfu}).
While formally subleading, it is well known that such
$\mathrm{N^{3}LL}_{\sigma}$ Sudakov terms can be large \cite{deFlorian:2001zd}.
Furthermore, besides $\mathrm{N^{3}LL}_{\sigma}$ ambiguities, also spurious,
finite, non-logarithmic
$\mathcal{O}(\alpha_{{\scriptscriptstyle \mathrm{S}}}^{2})$
terms in \STJ{} can contribute $\Phi$-dependent differences between
$d\sigma_{{\scriptscriptstyle \mathcal{M}}}/d\Phi$ and
$d\sigma_{{\scriptscriptstyle \mathrm{NLO}}}^{{\scriptscriptstyle \mathrm{ST}}}/d\Phi$.
Given these reasons, together with the fact that the primary objective
is to render the \STJ{} simulation NLO accurate for both $t$-channel
single-top and $t$-channel single-top plus jet observables, we allow
for a general dependence on $\Phi$ in the fitted $\mathcal{A}_{2}$ coefficient.

If our postulate is correct then the fitted quantity we obtain,
being of Sudakov origin, will be universal to $\mathrm{NNLL}_{\sigma}$
accuracy. Moreover, if the
leading $\mathcal{O}(\sqrt{\alpha_{{\scriptscriptstyle \mathrm{S}}}})$
ambiguity in that fitted coefficient is also completely due to a
deficiency in the \MINLO{} Sudakov, at $\mathrm{N^{3}LL}_{\sigma}$,
that too will be universal; more specifically, it should not depend on
the collider centre-of-mass energy or the PDFs.  That hypothesis has
support from the fact that, in all earlier work, the inclusion of
higher order terms in the Sudakov form factor has been all that was
required to promote \MINLO{} simulations of jet-associated production
processes to NLO accuracy for their inclusive
analogues~\cite{Hamilton:2012rf,%
  Luisoni:2013kna,Frederix:2015fyz,Hamilton:2016bfu}.  If true,
$\mathcal{A}_{2}$ fits performed for a given collider setup would
formally maintain the equality in
eq.~\eqref{eq:sect223-dsigMiNLO-eq-dsigRF-eq-dsigNLO} up to
$\mathcal{O}(\alpha_{{\scriptscriptstyle \mathrm{S}}}^{2})$ terms when
used in the context of other setups, e.g.~with different
centre-of-mass energies and/or different PDFs. So, while the fits
performed in this paper correspond to a specific 13 TeV LHC setup, we
expect that, in practice, they should be somewhat robust against
changes to it. This is exactly what we find when using the fits for
generating 8 TeV results, see
appendix~\ref{app:Tune-extrapolation}. We advise, however, that,
in general, if a fit is carried out for some given collision energy
and then used to reweight \STJ{} events simulated at a much higher
one, the enlarged phase space of the latter means that the neural
network model will need to interpolate outside the region covered
by the data used to train it, e.g. in regions with very high transverse
momentum/rapidity top quarks. Hence, we expect our
approach to work well provided the neural network Sudakov form factor
tuning is not applied when simulating \STJ{} production at energies
significantly above that used in its training. 


\section{\MINLO{} fit extrapolation from 13 TeV to 8 TeV}
\label{app:Tune-extrapolation}

\begin{figure}[t]
\begin{center}  
  \includegraphics[width=0.37\textwidth,height=0.36\textheight]{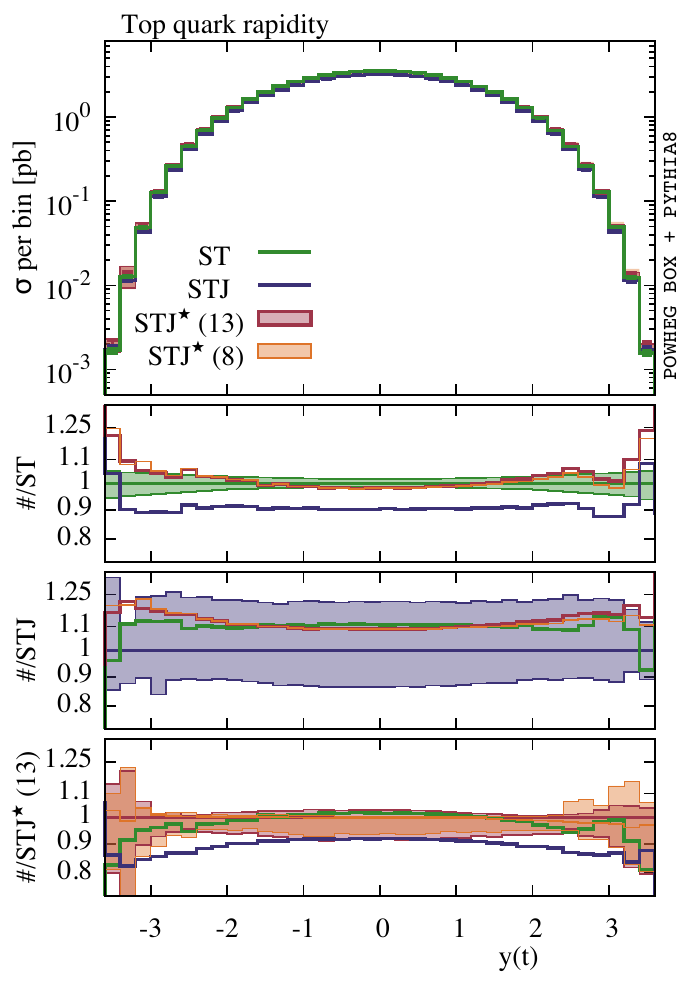}  
  \hspace{20 mm}
  \includegraphics[width=0.37\textwidth,height=0.36\textheight]{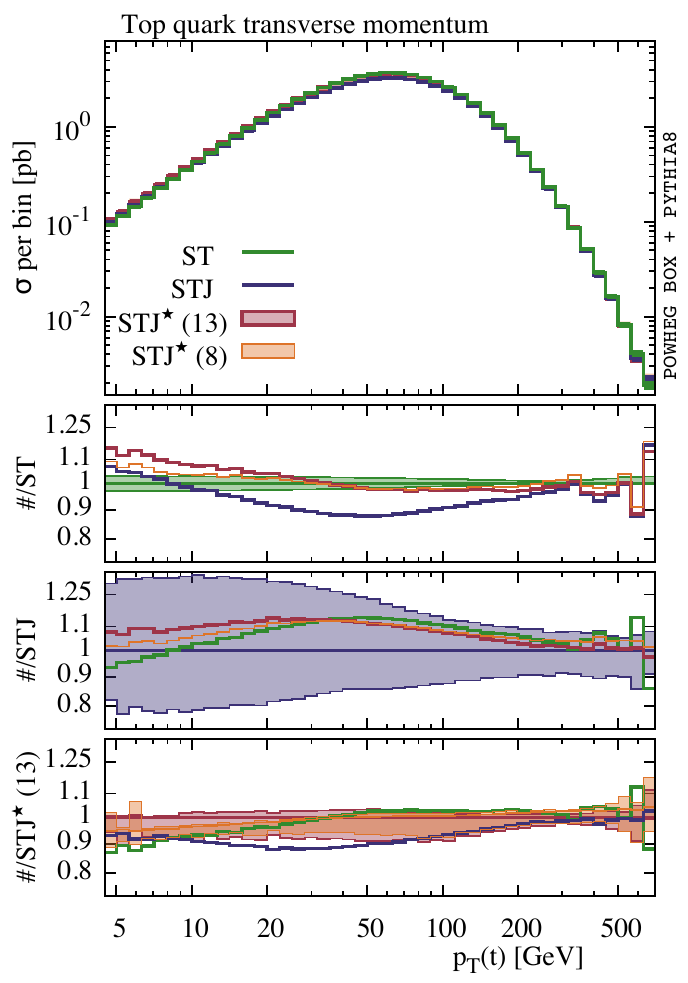}
\end{center}
\caption{
  Here we show the same predictions as in Fig.~\ref{fig:top}, but for
  an 8 TeV rather than a 13 TeV LHC.
  The left-hand plot shows the rapidity of the top quark in $t$-channel
  single-top production, while the right-hand plot shows its transverse
  momentum.
  As with the \STJR{} predictions in the main text, the \STJR{}\ (13)
  predictions (red) are obtained by tuning the \MINLO{}
  Sudakov form factor using 13 TeV LHC \ST{} and \STJ{} event samples.
  The \STJR{}\ (8) predictions (orange) are obtained by tuning the
  \MINLO{} Sudakov form factor using 8 TeV LHC \ST{} and \STJ{} event
  samples. The very good level of agreement between \STJR{}\ (13) and
  \STJR{}\ (8) predictions points to a high degree of universality in
  the Sudakov form factor corrections output by the tuning procedure.
} 
\label{fig:top_8tev}
\end{figure}

In this appendix we show a representative sub-sample of the
distributions presented in the main text for the 13 TeV LHC,
here, instead, for the 8 TeV LHC. The purpose of the presentation
is to give an indication as to how well the tuned \MINLO{}
Sudakov form factor in \STJR{}, obtained by carrying out the
neural network fitting procedure (section~\ref{sec:NN}) using
13 TeV \ST{} and \STJ{} events, can perform under different
running conditions to those used for training the network.

To further gauge the universality of the output of the Sudakov form
factor tuning procedure, the red and orange lines in the plots of this
subsection compare 8 TeV predictions obtained with a \STJR{}
simulation, tuned on 13 TeV event samples, (red), to those of a
\STJR{} simulation tuned on the same 8 TeV event samples (orange) used
to make the \ST{}, (green), and \STJ{}, (blue), results. In general we
observe a remarkable level of agreement in the predictions obtained
with the two different \STJR{} tunes.


%
%
\begin{figure}[htbp]
\begin{center}  
  \includegraphics[width=0.37\textwidth,height=0.33\textheight]%
                  {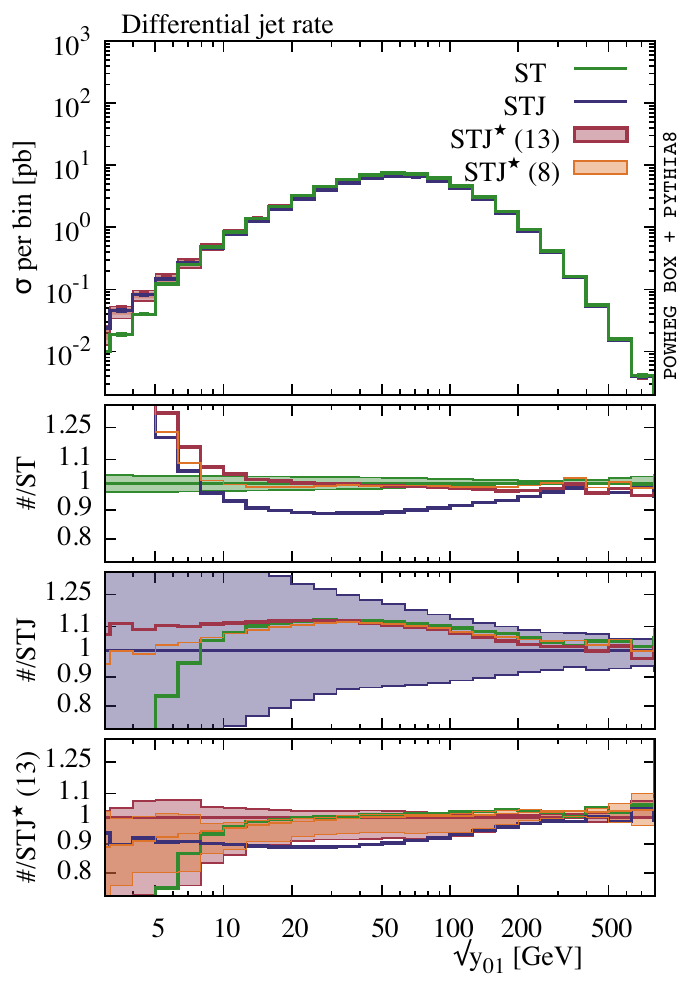}  
  \hspace{20 mm}
  \includegraphics[width=0.37\textwidth,height=0.33\textheight]%
                  {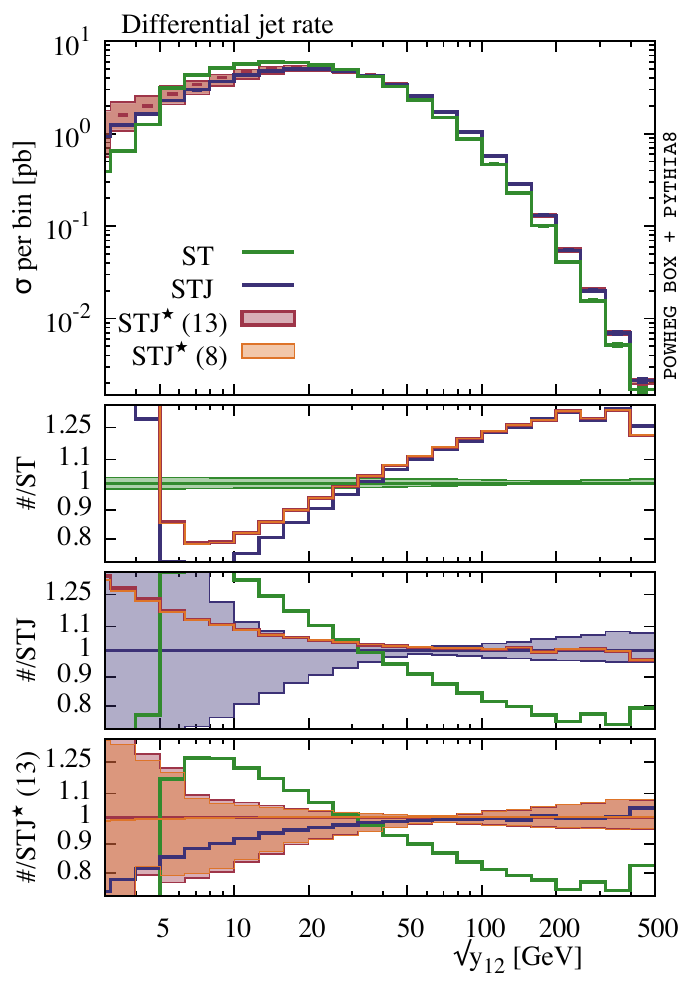}
\end{center}  
\caption{
  $0 \rightarrow 1$ (left) and $1 \rightarrow 2$ (right) differential
  jet rates in the $k_{t}$ clustering algorithm, at the 8 TeV LHC.
  As in Fig.~\ref{fig:top_8tev}, the \STJR{}\ (13) program (red) uses
  a \MINLO{} Sudakov form factor tuned using 13 TeV LHC \ST{} and \STJ{}
  samples, according to sections~\ref{subsec:Improved-STJ-MiNLO}-\ref{sec:NN}.
  The \STJR{}\ (8) program analogously uses a \MINLO{} Sudakov form
  factor tuned with 8 TeV LHC \ST{} and \STJ{} samples.
}
\label{fig:diffjetrates_8tev}
\end{figure}
\begin{figure}[htbp]
\begin{center}  
  \includegraphics[width=0.37\textwidth,height=0.33\textheight]%
                  {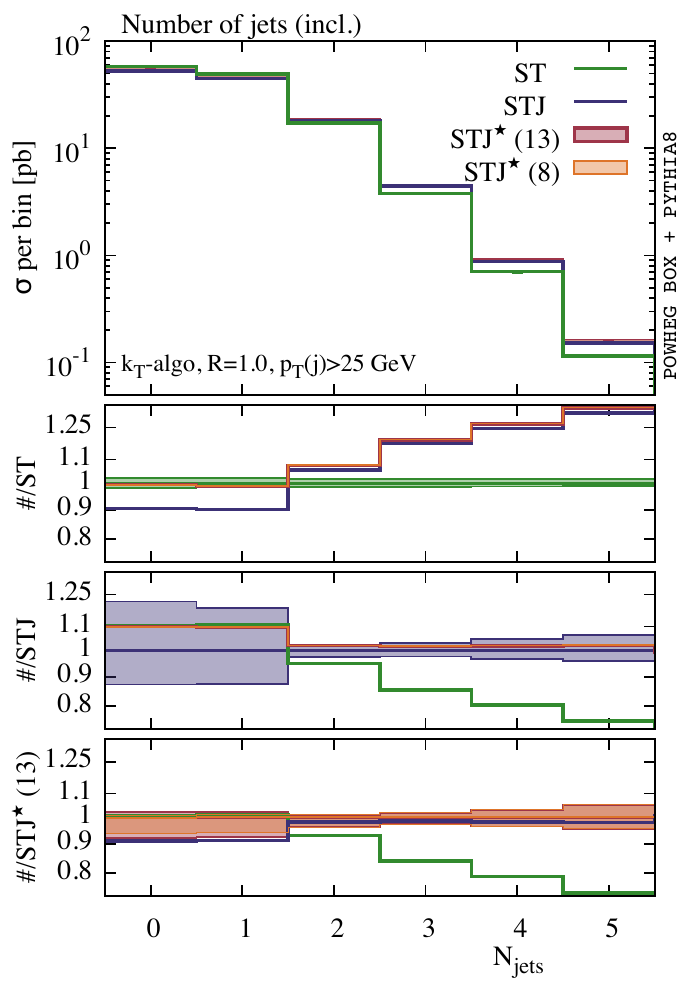}
  \hspace{20 mm}
  \includegraphics[width=0.37\textwidth,height=0.33\textheight]%
                  {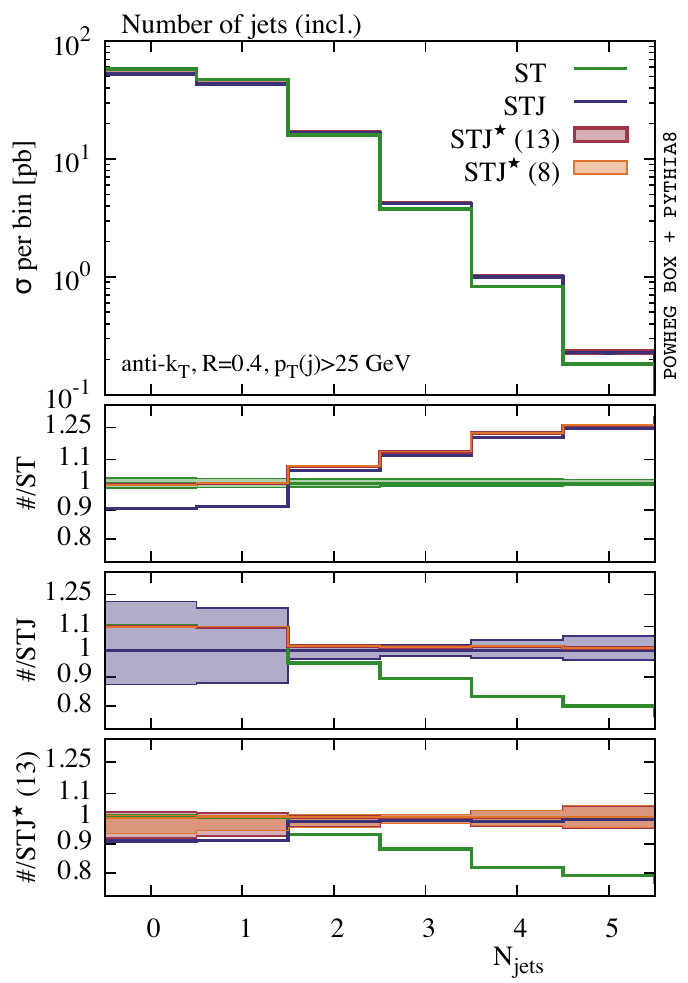}
\end{center}  
\caption{Inclusive jet cross sections at the 8 TeV LHC. The left-hand
  plot shows predictions for jets defined according to the $k_{t}$
  clustering algorithm with radius parameter $R = 1$. The right-hand
  plot gives predictions for the anti-$k_{t}$ algorithm with $R = 0.4$.
  The \STJR{}\ (13) program (red) uses a \MINLO{} Sudakov form factor
  tuned using 13 TeV LHC \ST{} and \STJ{} events
  (sections~\ref{subsec:Improved-STJ-MiNLO}-\ref{sec:NN}).
  The \STJR{}\ (8) program uses a \MINLO{} Sudakov form
  factor tuned with 8 TeV LHC \ST{} and \STJ{} events.
  \label{fig:sigincl_8tev}
  }
\end{figure}

\begin{figure}
\begin{center}  
  \includegraphics[width=0.37\textwidth,height=0.36\textheight]{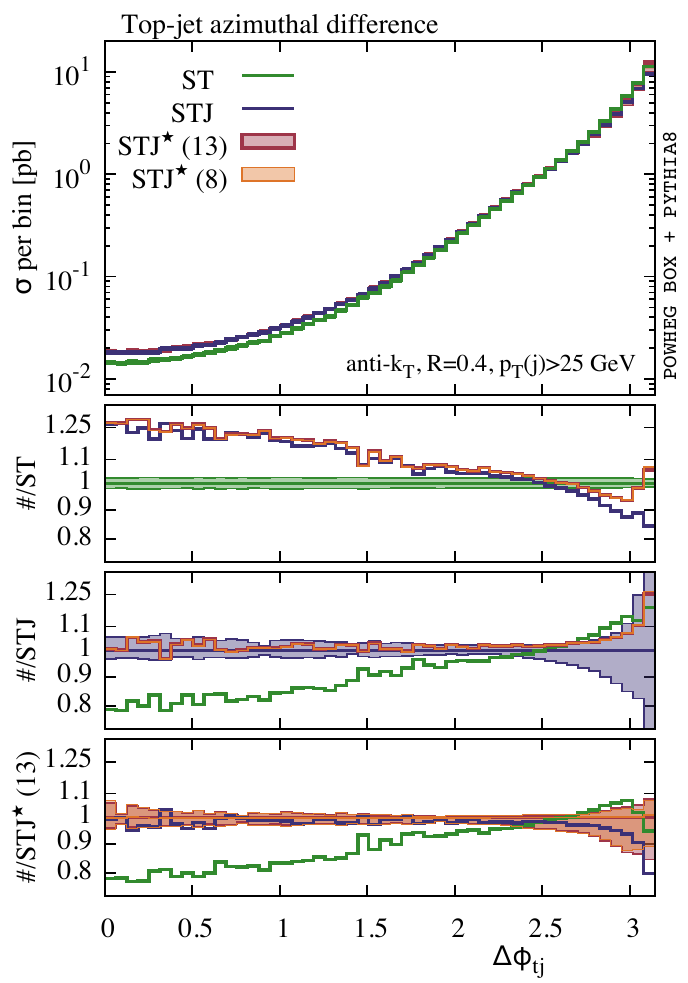}
  \hspace{20 mm}
  \includegraphics[width=0.37\textwidth,height=0.36\textheight]{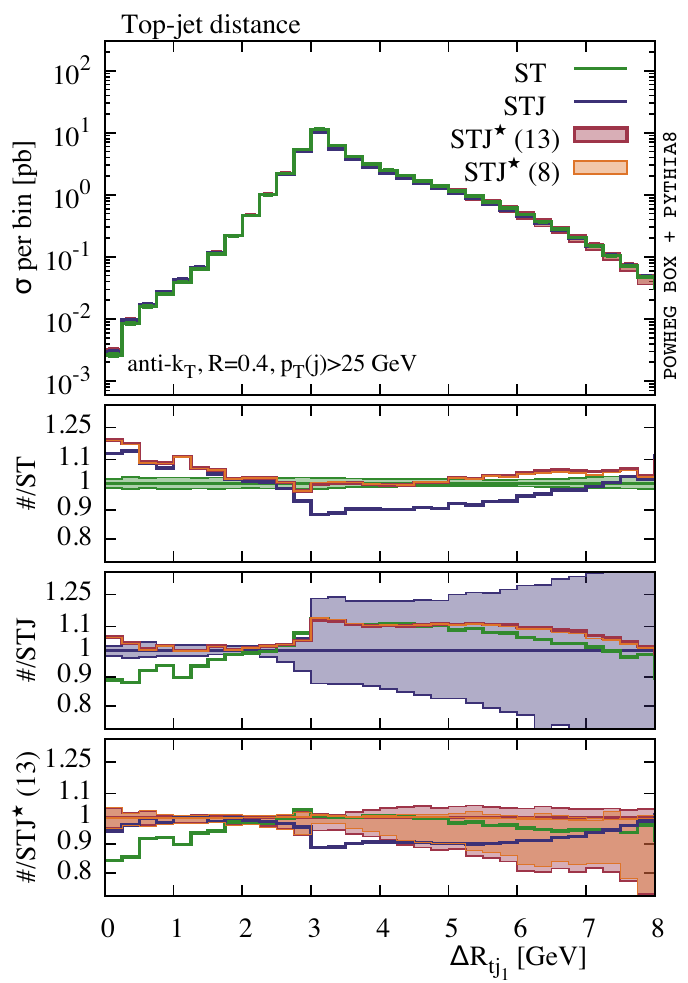}  
\end{center}  
\caption{
  The left-hand plot shows predictions for the azimuthal angle between
  the top quark and the leading jet in $t$-channel single-top production
  at the 8 TeV LHC. The right-hand plot similarly shows predictions for
  the distance between the same two objects in the $\eta-\phi$ plane.
  The colour coding and naming of the various predictions is as in
  Figs.~\ref{fig:top_8tev}-\ref{fig:sigincl_8tev}.
}
\label{fig:DRtj_8tev}
\end{figure}


\bibliographystyle{JHEP}
\bibliography{paper}

\end{document}